\documentclass[journal]{IEEEtran}
\IEEEoverridecommandlockouts

\usepackage{tabularx}
\usepackage{amsmath,amsfonts,amssymb,amscd}
\usepackage[pdftex]{graphicx}
\usepackage{cite}
\usepackage{booktabs}
\usepackage{paralist}
\usepackage{enumitem}
\usepackage[export]{adjustbox}
\usepackage{multicol}
\usepackage{fancyvrb}
\usepackage{steinmetz}
\usepackage{mathtools}
\usepackage{siunitx}
\usepackage{wrapfig}
\usepackage[caption=false]{subfig}
\usepackage{setspace}
\usepackage{multirow}
\usepackage{soul,xcolor} 

\usepackage{fancyhdr}
\makeatletter
\def\footnoterule{\kern-3\p@
  \hrule \@width 2in \kern 2.6\p@} 
\makeatother

\graphicspath{{./figures/}}

\newcommand{\red}[1]{\textcolor{red}{\st{#1}}}
\newcommand{\blue}[1]{\textcolor{blue}{{#1}}}
\renewcommand{\red}[1]{}
\renewcommand{\blue}[1]{#1}

\begin{document}

\title{\red{High Voltage Generation by Fiber Optic Power \\} \blue{High Voltage Generation by Fiber-Coupled Pulsed Laser for a Simple Receiver Circuit Structure}
\thanks{This material is based upon work supported by the National Science Foundation under Grant No. 1808489.}
}


\author{
    \IEEEauthorblockN{Sanghyeon Park and Juan Rivas-Davila\\
    Stanford University\\
    \{spark15, jmrivas\}@stanford.edu}
}

\maketitle
\thispagestyle{fancy}

\begin{abstract}
Almost all high-voltage dc generation for low-power applications is done by either electrostatic machines or voltage multipliers.
Electrostatic machines use mechanically moving parts to transfer charge and energy from the low-voltage side to the high-voltage side.
Voltage multipliers use capacitive and inductive networks to achieve the same purpose of energy transfer.
Considering the pros and cons inherent in those mechanical, capacitive, and inductive energy transfer, a new means of energy transfer may provide a superior design of a high voltage dc generator.
Here we investigate the design of high voltage generators based on optical power transfer.
Optical power delivered via fiber-optic cable allows extensive input-to-output dc insulation and spatial separation.
These characteristics lead to advantages with respect to ease of insulation, ease of electromagnetic shielding, and scalability.
We experimentally validate the idea by building and testing a 5.5~kV dc generator module solely powered by a 20~kHz pulsed laser, and cascading three of those modules to obtain 14.7~kV dc output voltage.
We then discuss possible improvements to the circuit design to make it useful for real-world applications.
\blue{Finally, we demonstrate an optically powered electroadhesion gripper to show the practicality of the proposed high voltage generator.}

\end{abstract}

\IEEEpeerreviewmaketitle

\section{Introduction}
\label{sec:intro}

High voltage power supplies with an output in the kilovolts range are essential for numerous scientific instruments that need high voltage dc bias.
Examples include avalanche photo diodes (500~V) \cite{bartoloni2001power}, mass spectrometers (10~kV and higher) \cite{bajic2006mass}, and various photon and particle detectors for spacecrafts (from 150~V to 26~kV) \cite{bever2006high}.
Most of those applications draw only tens of microamperes of current or even less from the power supply.
Such an extremely low power requirement enables some unique ways to generate the necessary dc bias voltage.
This paper explores a high voltage dc generation technique using a pulsed laser delivered through a fiber optic cable.

A voltage multiplier is arguably the most popular technique to generate a high voltage dc, appearing in many recent papers~\cite{Young2013,Muller2016,Xu2018}.
The multiplier consists of two major parts.
The first is a number of cascaded rectifiers made of diodes and capacitors.
The other is a mechanism to block dc voltage and pass ac voltage from the input terminal to each of the rectifiers.
There is little variation in the rectifier structure; 
it always consists of many half-wave rectifiers (two serialized diodes) connected in series, and sometimes in parallel as well to accommodate multi-phase inputs.
On the other hand, the dc-blocking mechanism exhibits a great variation with different types of components and their arrangements.

The dc-blocking ac-coupling devices in multipliers are mostly either a capacitor network or a combination of capacitor and transformer networks.
The capacitor network realizes capacitive coupling between the input terminal and rectifiers by presenting a low impedance at the frequency of operation.
Famous topologies based on capacitor networks include Cockcroft-Walton~\cite{greinacher1921methode,cockcroft1932experiments} and Dickson~\cite{cleland1960new,dickson1976chip}.
The transformer network realizes inductive coupling by using multi-winding transformers of high interwinding breakdown voltage.
Examples of using the combination of both capacitor and transformer networks are a 300~kV generator developed by Enge~\cite{enge1971cascade}, a 160~kV generator by Mao \textit{et al.}~\cite{8356118}, and a 100~kV generator by Pokryvailo \textit{et al.}~\cite{pokryvailo2010high}.

Interestingly, the task of blocking dc and coupling ac through a high-voltage insulation barrier is similar to transferring power across a distance, commonly known as wireless power transfer (WPT).
Capacitive and inductive couplings are indeed two most common approaches for near-field (nonradiative) power transfer.
But there is another category of WPT techniques called far-field (radiative) power transfer which uses visible light and microwaves as a means of power transmission.
Considering those parallels between high voltage dc generation and WPT, it is natural to ask whether the idea of optically generating a high voltage is useful, and if so, how.

Several studies, mostly from the field of power management integrated circuits, have shown the feasibility of high voltage generation by optical power transfer~\cite{lee1995miniaturized,ortega2008high,rentmeister2020120}.
Lee \textit{et al.}~\cite{lee1995miniaturized} achieved an open-circuit voltage of 150~V using an integrated circuit of 100 series-connected photovoltaic (PV) cells.
Similarly, Ortega \textit{et al.}~\cite{ortega2008high} achieved an open-circuit voltage of 103~V using 169 PV cells.
Rentmeister \textit{et al.}~\cite{rentmeister2020120} used 196 series-connected PV cells to produce a 125~V open-circuit voltage.

Our work differs from the existing literature by not relying on a massive number of PV cells for high output voltage.
A resonant operation combined with a step-up transformer provides enough voltage gain, allowing us to build a high voltage generator using only low-cost commercially available parts and assembly service.
Considering the often prohibitive cost of custom-built high voltage integrated circuits, our work is attractive in that it makes available optically powered high-voltage generators to a wider range of research projects and applications.

In this paper, we firstly review existing techniques for high voltage dc generation in section~\ref{sec:review}.
Then we discuss viable configurations of light-based high voltage generators in section~\ref{sec:optical}.
The purpose of these two sections is to put the proposed light-based scheme into perspective by comparing it with other generator designs of different working principles.
We then proceed to implement and test the proposed voltage multiplier, of which the experimental results are given in section~\ref{sec:prototype}.
Also presented in the section are the practical limits we experienced during our design process and a projection on achievable performance.
\blue{To demonstrate the practicality of the proposed system, we show in section~\ref{sec:app_example} an application of it to an electroadhesive gripping device.}
We conclude the paper in section~\ref{sec:conc} with a couple of questions worth exploring for future research.

\section{Review of Popular Techniques for High Voltage Generation}
\label{sec:review}


This section reviews some of the most popular types of high voltage dc generators.
Designs are grouped based on how the input power from the low-voltage side is transferred through an insulation barrier to the high-voltage side.
Their working principles and associated pros and cons make each of the designs suitable for different applications with varying degrees of power, noise, size, space, and reliability requirements.

\subsection{Mechanical coupling}

The first to be discussed is a group of devices in which the power is conveyed mechanically from low- to high-voltage sides.
Commonly called \textit{electrostatic generators}, those devices are the first to appear in the history of high voltage dc generators, dating back as far as 1663~\cite{de1950guericke}.
Just to name some of the most famous designs we refer to Kelvin water dropper~\cite{thomson1868xvi}, Wimshurst influence machine~\cite{wimshurst1893xxiv}, Van de Graaff generator~\cite{van1936design}, Felici's rotating cylinder~\cite{felici1962rotating}, and a varying capacitance machine~\cite{6443559,4080409}.

\begin{figure}[b]
     \centering
     \includegraphics[width=\linewidth]{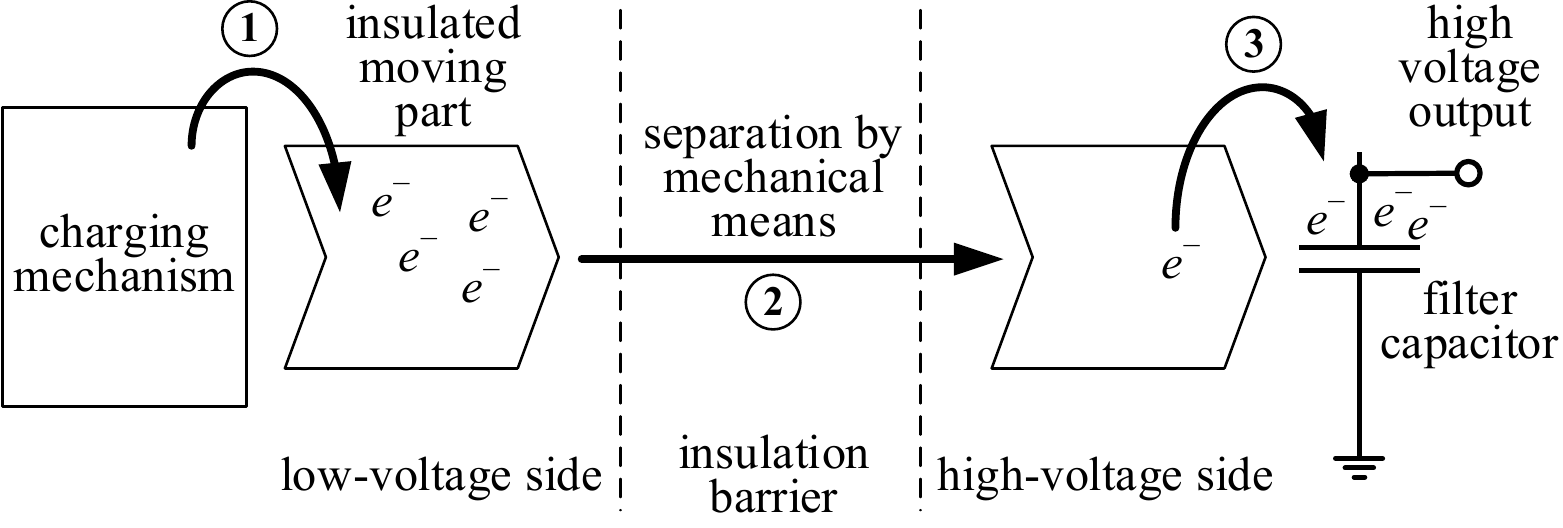}
     \caption{The working principle of electrostatic generators, which generate a high-voltage dc by using mechanical energy coupling.}
     \label{fig:mechanical}
\end{figure}

Fig.~\ref{fig:mechanical} depicts the working principle of electrostatic generators.
Firstly, an insulated object is loaded with some amount of static charge (marked with \textcircled{\small 1} in Fig.~\ref{fig:mechanical}).
Methods for charging the insulated object include the triboelectric effect, the electrostatic induction, an electret, a corona discharge, and a direct contact with a voltage source such as a battery.

Secondly, the object is moved across the insulation barrier toward the high-voltage side mechanically (marked with \textcircled{\small 2}).
As the object is pulled away from the low-voltage side, the mechanical energy is converted into electric potential energy of the charge.
Popular methods to actuate the moving part include a rotor, a circulating belt, and a vibrator.

Lastly, after the charge has gained enough electric potential energy, it is dumped to a filter capacitor (marked with \textcircled{\small 3}).
The role of the capacitor is to smooth the output voltage ripple and, in the case of pulsed applications, to accumulate enough energy for a high-current discharge.
Once this step is completed, the moving part is returned to the low-voltage side and the process repeats.


After more than a century of study, electrostatic generator is still a subject of active research.
This largely owes to the advances in the field of micro-electromechanical systems (MEMS) and the microfluidics technology, combined with the interest in developing a miniature energy harvesting generator to power small portable devices.
Recent works based on microfluidics technology include a chip-based Kelvin dropper~\cite{marin2013microfluidic}, a ballistic Kelvin dropper~\cite{xie2014pressure}, and a mercury-droplet-based influence machine~\cite{conner2018energy}.
Recent works based on MEMS technology include a varying capacitance machine~\cite{mitcheson2004mems,miao2006mems} and an influence machine~\cite{le2016mems},

The last to mention is a peeling-tape-based X-ray source\cite{camara2008correlation,putterman2014mechanoluminescent}.
The operation is based on triboluminescence, a phenomenon that can be understood by the same principle discussed in this section:
the peeling of the tape creates and separates a static charge (\textcircled{\small 1} and \textcircled{\small 2} of Fig.~\ref{fig:mechanical}, respectively);
this charge immediately jumps across the vacuum insulation gap and its high-energy recombination generates X-ray.


\subsection{Capacitive and inductive coupling}

\begin{figure}[b]
     \centering
     \includegraphics[width=0.77\linewidth]{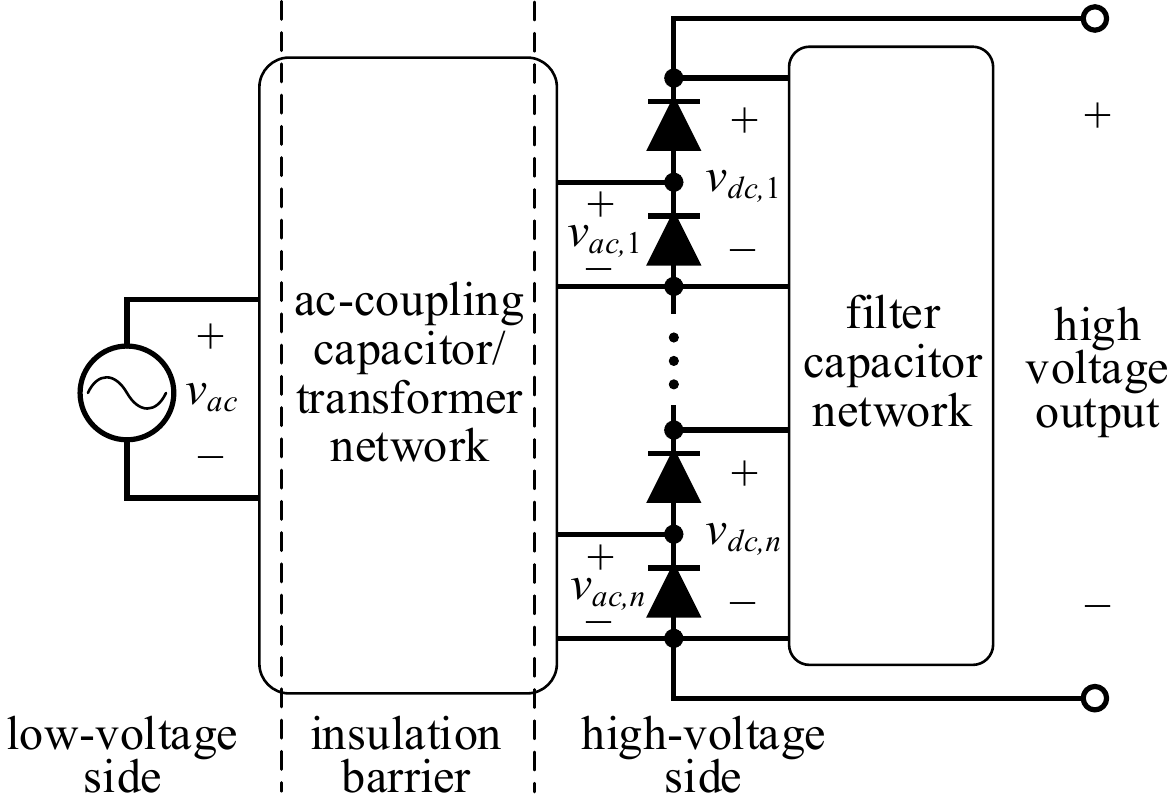}
     \caption{The structure of voltage multipliers, which transfer the energy from low- to high-voltage side via capacitive and/or inductive coupling.}
     \label{fig:capacitive_inductive}
\end{figure}

Capacitive and inductive coupling is extensively used for building a high voltage dc generator.
Fig.~\ref{fig:capacitive_inductive} shows a generalized structure of such system.
It consists of an ac voltage source, a capacitive and/or transformer network, an $n$-stage voltage multiplier (i.e., $n$ half-wave rectifiers in series), and a filter capacitor network.

The \textit{ac-coupling network}'s role is to transfer the ac voltage $v_{ac}$ from the low-voltage side to the high-voltage side through the insulation barrier.
Ideally, the coupling network presents zero impedance at the switching frequency, thus the ac voltages $v_{ac,1}, \dots, v_{ac,n}$ are equal in amplitude to the input ac voltage $v_{ac}$.
The coupling network may be built with capacitors (e.g., \cite{greinacher1921methode,cockcroft1932experiments,cleland1960new,dickson1976chip}), transformers (e.g., \cite{hull1941electric,harding1949voltage,thompson1989magnetic}), or both (e.g., \cite{enge1971cascade,8356118,pokryvailo2010high}).
A step-up transformer with a Cockcroft-Walton multiplier is a popular combination and often appear in literature.
Different capacitor-transformer structures result in different dc voltage stress that each of the components need to withstand, their necessary capacitance and inductance values, as well as the output impedance and power of the multiplier.
The specifics of the coupling network's structure are determined based on the required performance, limits on the total space and weight, availability of high voltage parts, etc.

The \textit{filter capacitor network} provides dc voltages denoted by $v_{dc,1}, \dots, v_{dc,n}$ by smoothing the half-wave rectifiers' output.
The capacitors in the network are connected in either series or parallel, or sometimes a mixture of both.
Because a voltage multiplier is often built with only capacitors and diodes, many multiplier topologies are named based on their capacitor network structures.
Cockcroft-Walton (or Greinacher) type~\cite{cockcroft1932experiments,greinacher1921methode} is when all the capacitors are serialized in both the ac-coupling capacitor network and the filter capacitor network.
Dickson type~\cite{cleland1960new,dickson1976chip} is when all the capacitors are in parallel.
A mixture of series and parallel capacitors can also be found in literature, for example, \cite{6497641,8767979}.

\section{High Voltage by Optical Power Transfer}
\label{sec:optical}

Compared to high voltage generators based on mechanical coupling (electrostatic generators) and capacitive-inductive coupling (voltage multipliers), investigations on optical power transfer for high voltage generation are relatively lacking.
Optical power-based systems have many potential benefits which are to be discussed in this section.
The biggest downside of optical power transfer is the low power efficiency and consequently a low available output power.
Therefore, this scheme should be used only for applications that need high voltage but not much power.

Here, we focus on a system that uses fiber-coupled laser to transfer power.
The reason is twofold:
First, a laser of sufficiently high output power and high intensity is readily available on the market.
Second, using a fiber-coupled laser allows us to configure the experiment in a flexible manner without having to fix the relative location of the light source and the receiver.

\subsection{System powered by a continuous-wave laser}

\begin{figure}
     \centering
     \includegraphics[width=\linewidth]{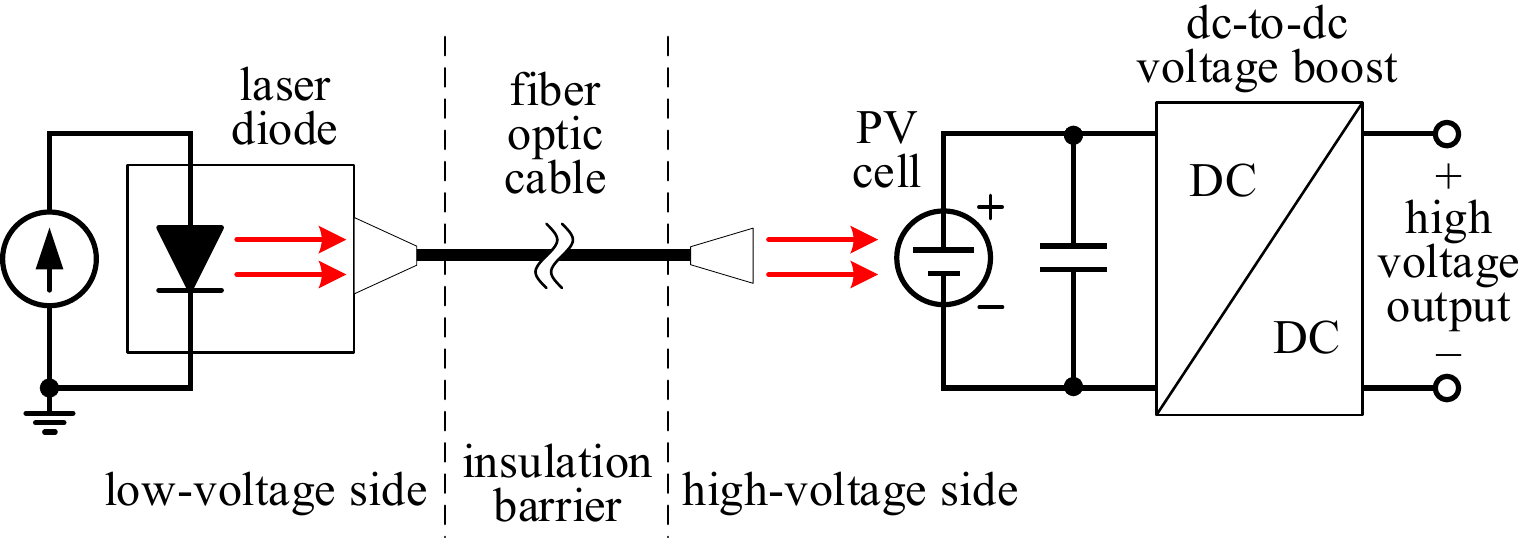}
     \caption{The configuration of a high voltage generator based on fiber-coupled continuous-wave laser.}
     \label{fig:config_cw}
\end{figure}

Using fiber-coupled laser to deliver power to low-voltage electronics has been investigated by several authors~\cite{werthen2008powering,wake2008optically,faria2008method,furey2010power}.
One can easily conceive an approach where the load electronics in those works are replaced with a dc-to-dc high voltage generator.
Fig.~\ref{fig:config_cw} shows the configuration of such system.
Using continuous-wave laser allows one to operate the PV cell at its maximum power point.
This potentially leads to a high efficiency as well as high available power.


If a sufficiently large number of PV cells are in series, the voltage produced from the PV array can be high enough such that the dc-to-dc voltage boost is not necessary.
Such is the case in works by Lee \textit{et al.}~\cite{lee1995miniaturized} and Ortega \textit{et al.}~\cite{ortega2008high} where authors integrated over a hundred of PV cells on a silicon die to produce an open-circuit voltage of 100~V and higher.

The work by Rentmeister \textit{et al.}~\cite{rentmeister2020120} is perhaps the closest to the strategy described in Fig.~\ref{fig:config_cw}.
The authors firstly produced 125~V using 196 on-chip PV cells, then used an off-chip Dickson voltage multiplier to generate over 600~V dc output.
The authors made it clear that their method can achieve 1.5~kV output voltage and even higher by simply increasing the number of PV cells in the array.

One disadvantage of a continuous-wave based high-voltage generation is the necessity for an dc-to-ac converter (inverter).
Unless special care is taken such as implementing the inverter circuit on a low-power integrated circuit~\cite{rentmeister2020120}, the dc-to-ac conversion may overburden the limited power budget of a solar cell.

Also, using a PV array necessitates balancing the optical power among individual cells so as to maximize the output power from the array.
Spreading the light evenly over multiple cells may add to the difficulty of implementation, especially when the light is from a fiber-optic cable which is effectively a point light source.
Moreover, such a massively integrated PV array is not readily available on the market; at this moment it is only available to those who can afford the time and cost for development of a custom integrated circuit.

If instead a single PV cell is used, the output voltage from the cell is around 0.6~V.
This voltage is insufficient for many electronic circuits and power transistors which demand at least 0.7~V and often higher.
There are some relaxation oscillators and integrated circuit solutions that work on a supply voltage of 0.6~V or less, for example, Joule thief oscillator~\cite{strauch2018restart} and an energy harvesting chip~\cite{ datasheet2013linear}.
However, they still suffer from a too low available power or output voltage.

\subsection{System powered by a pulsed laser}

\begin{figure}
     \centering
     \subfloat[][]{\includegraphics[width=\linewidth]{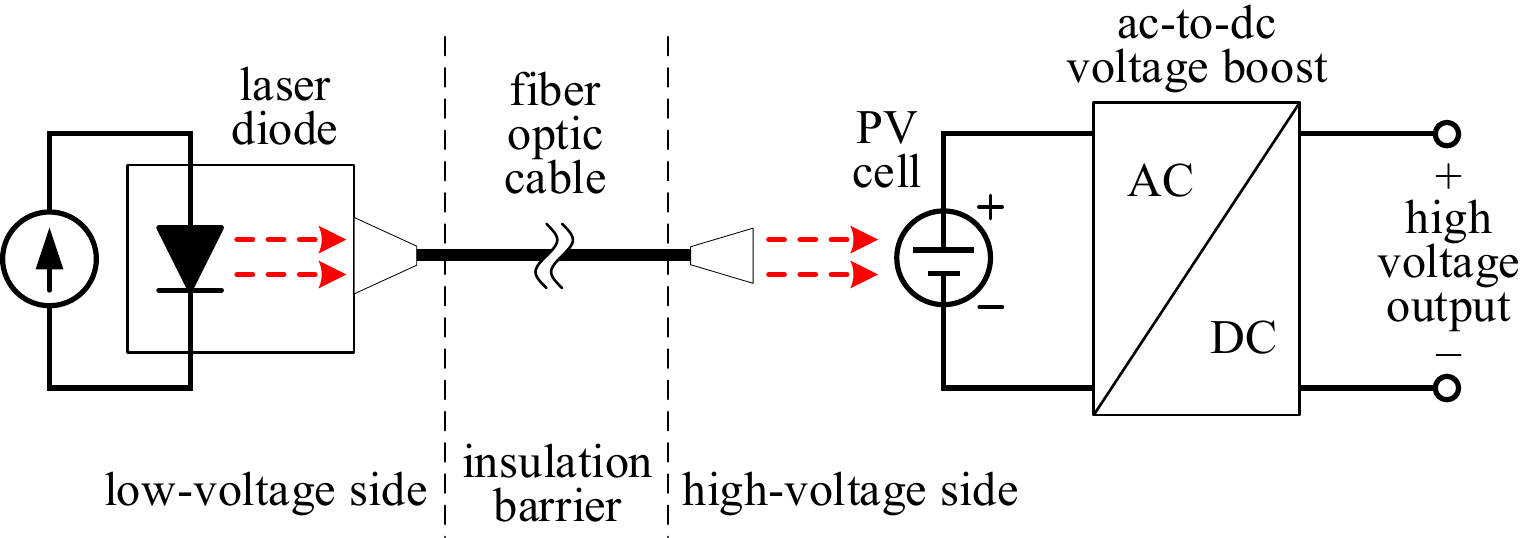}\label{fig:config_pulsed_sch}}
     \hfill
     \subfloat[][]{\includegraphics[width=0.83\linewidth]{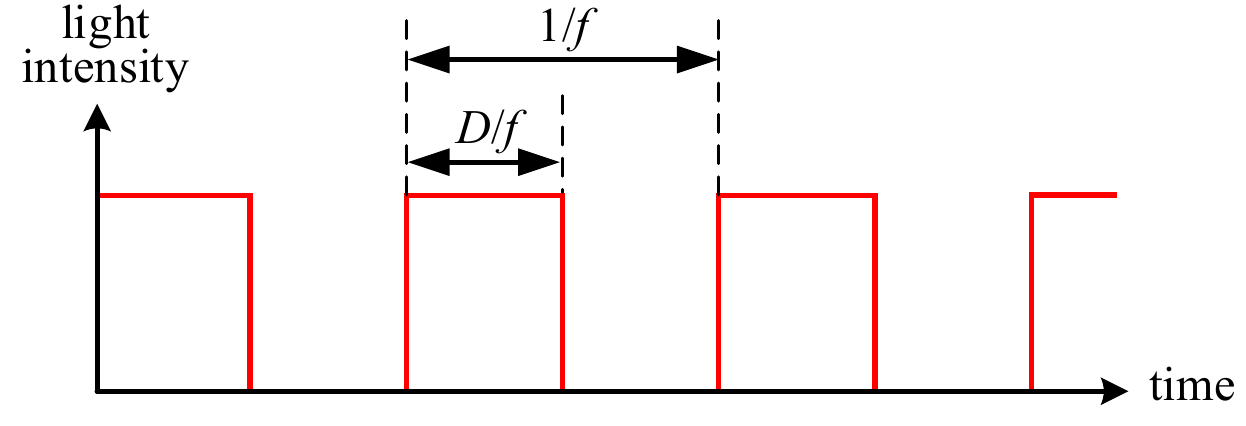}\label{fig:config_pulsed_plt}}
     \caption{The configuration of a high voltage generator based on fiber-coupled pulsed laser. (a) Schematic. (b) Light intensity versus time plot. $f$ and $D$ denote the switching frequency and duty cycle of the pulsed laser, respectively.}
     \label{fig:config_pulsed}
\end{figure}

In order to overcome the discussed shortcomings of a system based on continuous-wave laser, we propose a scheme based on a pulsed laser.
Fig.~\ref{fig:config_pulsed} illustrates the proposed arrangement.
The laser diode is driven on and off repeatedly with a set frequency $f$ and a duty cycle $D$ (Fig.~\ref{fig:config_pulsed_plt}) and consequently an ac voltage is produced from the PV cell.

The advantages of the proposed approach is as follows.
First, the receiving end does not need an inverter and thus the circuit design is simplified.
Whereas a continuous-wave system needs to perform dc-to-ac conversion followed by ac-to-dc conversion, the proposed pulsed system's receiving end only needs to perform ac-to-dc conversion.
As a result, any overhead associated with the inverter circuit is avoided.

Second, when multiple high-voltage circuits are cascaded, it is easy to stagger the phases of pulsed lasers driving multiple modules and have the ripples cancel each other.
Ideally, when $n$ modules are cascaded, one can reduce the ripple ratio by $n$ times \blue{or more} using this phase interleaving technique.
Previous works~\cite{baumann1983x,pokryvailo2010high,pokryvailo2015100,katzir2015split} have investigated the idea of reducing the output voltage ripple by segmenting a voltage multiplier and driving each segment at staggered phases.

\begin{figure}
     \centering
     \includegraphics[width=\linewidth]{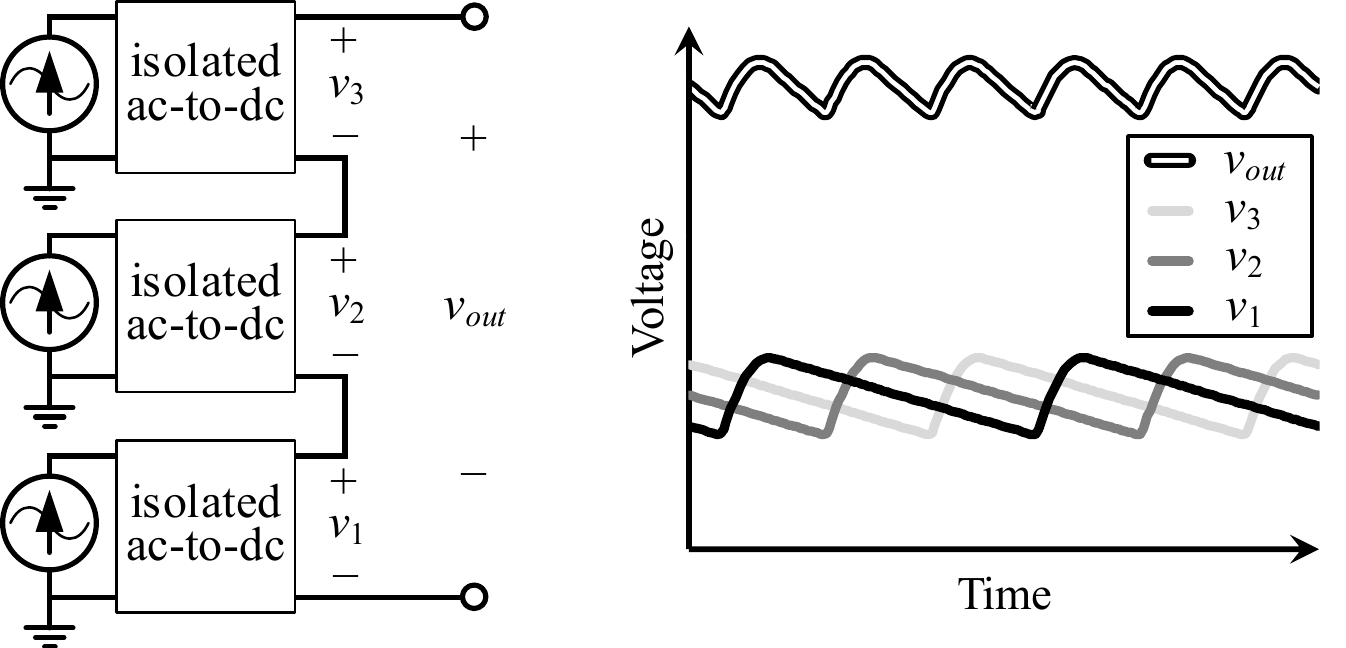}
     \caption{\blue{Conceptual illustration of ripple reduction by phase interleaving. Segmenting the circuit into three parts and driving them at 120$^{\circ}$ out of phase reduces the ripple amplitude of $v_{out}$ by more than three times.}}
     \label{fig:conceptual_ripple}
\end{figure}

\blue{Fig.~\ref{fig:conceptual_ripple} illustrates a conceptual ripple reduction when a multiplier is segmented into three parts and their switching phases are staggered by 120$^{\circ}$ to each other.
The output voltage $v_{out}$ is the sum of each ac-to-dc circuit's output voltages $v_1$, $v_2$, and $v_3$.
Compared to the case without interleaving where $v_1$, $v_2$, and $v_3$ waveforms overlap, the phase-interleaved scheme reduces the ripple amplitude of $v_{out}$ by more than threefold.
A more general in-depth study of the interleaving principle can be found in \cite{4126803}.
}

The proposed pulsed-laser system \red{realizes this idea with ease} \blue{can easily achieve the ripple reduction by phase interleaving} because all the laser diode drivers are placed on the low-voltage side and share the same ground potential.
In contrast, a high voltage generator consisting of continuous-wave-driven modules need means of communication between modules, otherwise the peak-to-peak ripple voltage will be that of the worst possible case because of frequency mismatches between oscillators.

\red{
Lastly, single- and few-cell PV modules producing a low voltage are readily available off-the-shelf thanks to the huge market size.
}

At the same time, this high voltage generation method has several limitations.
The first is a relatively low power efficiency, because the PV cell cannot operate at its maximum power point like the continuous-wave system.
The second is a relatively low power level, partly because of the low efficiency and partly because the power is transferred only for fraction of time.
Another limitation of the proposed method is that the ac voltage amplitude from the PV cell is inherently restricted to the turn-on voltage of the intrinsic p-n junction diode.

\subsection{Comparison with other power transfer methods}

\red{
In the context of our discussion, a light-based high voltage generator denotes a system which delivers power through the insulation barrier using a beam of light such as laser.
This scheme 
}

\blue{
Energy coupling through an insulation barrier by a pulsed laser
}
enables extensive electrical and physical isolation between the low-voltage input side and the high-voltage output side.
This feature leads to following advantages that do not exist in systems that rely on mechanical or capacitor-inductive energy couplings \blue{only}.

\red{
First, voltage multipliers can be cascaded as many as desired for a higher output voltage provided that adequate measures are in place to prevent corona discharge and breakdown.
In contrast, multipliers solely based on capacitor networks suffer a significant output voltage drop when too many rectifiers are in series.
To illustrate the point, for a capacitor-based voltage multiplier with $n$ rectifiers in series, the deviation of the actual output voltage from the ideal value is almost proportional to $n^3$~[37]--[39].
}

\blue{
First, optical energy coupling can be used to break down a long voltage multiplier into smaller segments and thus prevent an excessive output voltage drop.
For a conventional $n$-stage Cockcroft-Walton voltage multiplier, the deviation of the actual output voltage from the ideal value is often called an output voltage drop and denoted by $\Delta V$.
This $\Delta V$ is almost proportional to $n^3$.
By breaking down the $n$-stage multiplier into $m$ segments of $(n/m)$-stage multipliers, and powering each of those multipliers via optical power transfer, one can reduce $\Delta V$ by almost $m^2$ times.
}

\blue{
More specifically, the formula for $\Delta V$ of an $n$-stage Cockcroft-Walton multiplier is
\begin{equation}
    \Delta V = \frac{i_{out}}{fC}\left[\frac{2}{3}n^3 + \frac{1}{2}n^2 - \frac{1}{6}n\right]
    \label{equ:DeltaV}
\end{equation}
where $i_{out}$ is the average output dc current, $f$ is the ac input frequency, and $C$ is the capacitance value of coupling capacitors.
(The derivation of this formula can be found in many academic papers, e.g., \cite{cockcroft1932experiments,weiner1969analysis} and textbooks~\cite{kuffel2000high,Naidu2010,Ray2013}.)
If this multiplier is broken into $m$ segments in series, the new voltage drop $\Delta V'$ is (assuming $m$ is a factor of $n$)
\begin{align}
    \Delta V' & = \frac{i_{out}}{fC}\left[\frac{2}{3}\left(\frac{n}{m}\right)^3 + \frac{1}{2}\left(\frac{n}{m}\right)^2 - \frac{1}{6}\left(\frac{n}{m}\right)\right] \times m\\
	& = \frac{i_{out}}{fC}\left[\frac{2}{3}\left(\frac{n^3}{m^2}\right) + \frac{1}{2}\left(\frac{n^2}{m}\right) - \frac{1}{6}n\right].
    \label{equ:DeltaV_prime}
\end{align}
When $n$ is sufficiently large, $\Delta V'$ is almost $m^2$ times smaller than $\Delta V$.
}

The second advantage of building a multiplier based on optical power transfer is that it is easy to separate the high voltage part physically from the rest of the system.
Power from a fiber-coupled laser is easily routed losslessly via long and pliable fiber optic cables.
As a result, rectifiers on the receiving end of the cable can be arranged in any way the designer sees fit.
The physical and electrical isolation of the high voltage part can be taken advantage for better protection of low voltage circuits against arc discharge events.

The spatial separation also allows better protection of the load side from electromagnetic noise, which is important for sensitive scientific applications that need a very low noise environment.
In contrast, voltage multipliers based on transformer networks need careful design and placement of magnetic cores to minimize flux leakage.
Also, because magnetic cores are often rigid, brittle, and difficult to machine, they impose constraints on the placements of transformers and the circuits surrounding them.

\section{Prototype Circuit Driven by Pulsed Laser}
\label{sec:prototype}

\begin{figure}
     \centering
     \subfloat[][]{\includegraphics[width=\linewidth]{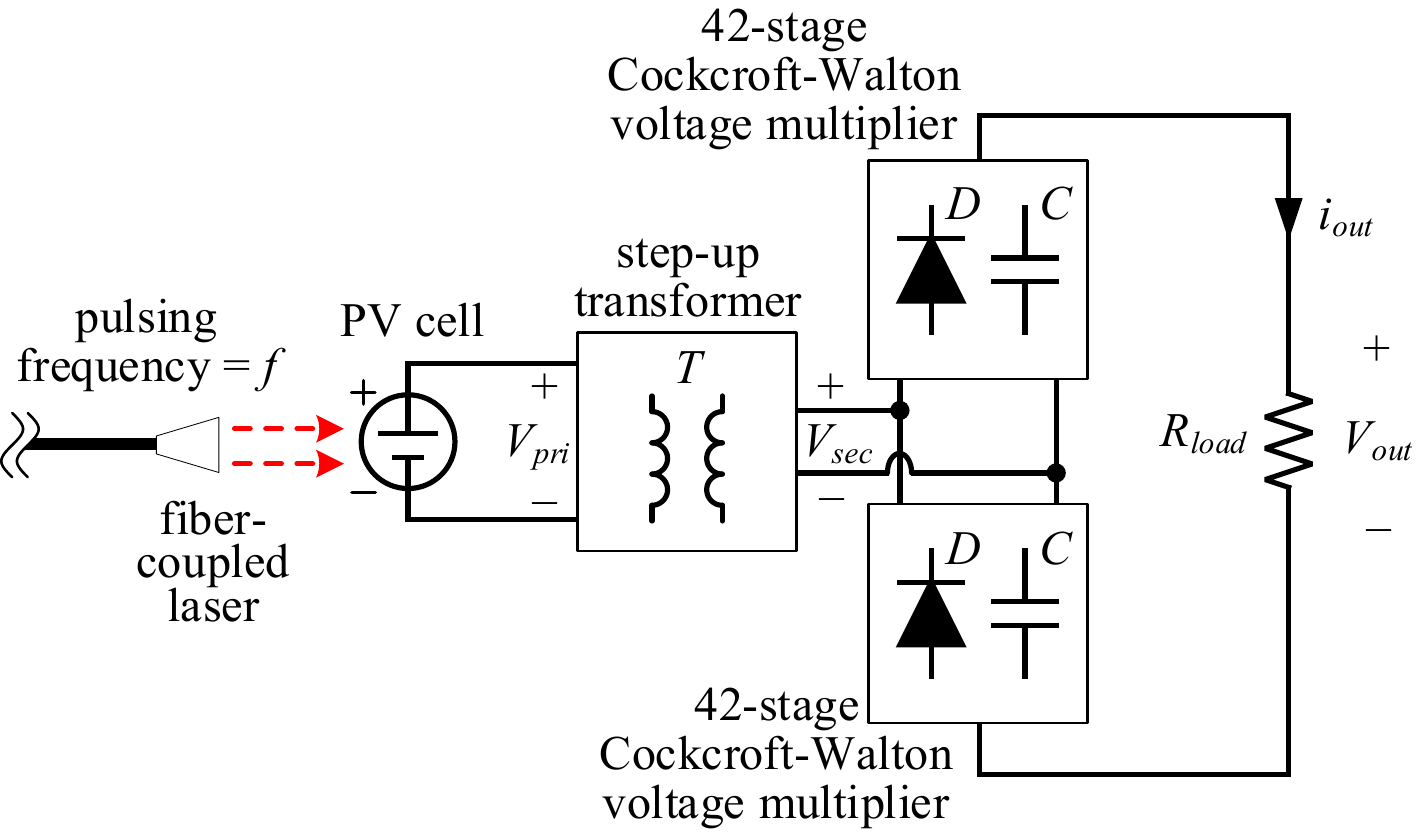}\label{fig:proto1_sch}}
     \hfill
     \subfloat[][]{\includegraphics[width=0.49\linewidth]{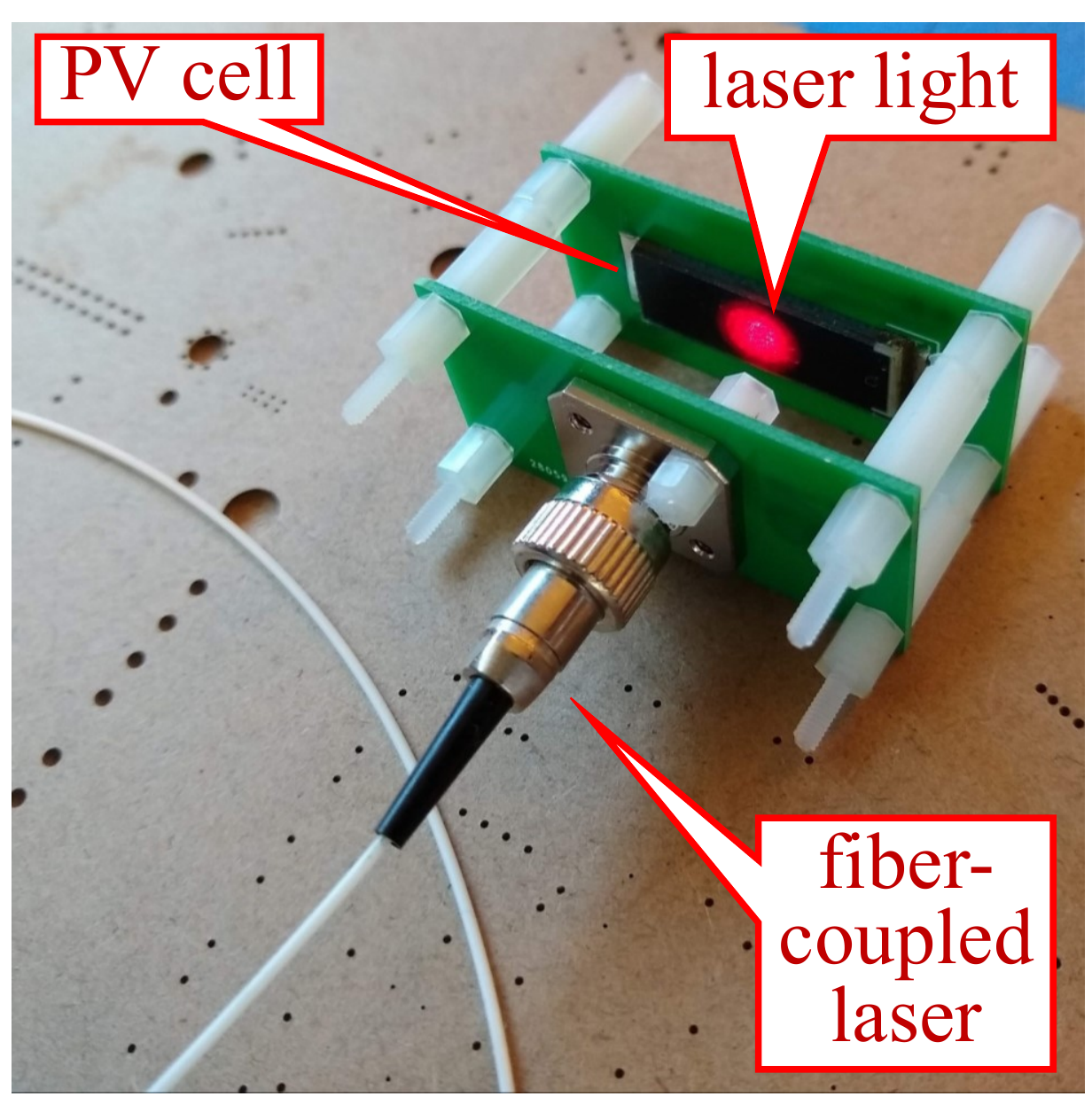}\label{fig:proto1_photo1}}
     \hfill
     \subfloat[][]{\includegraphics[width=0.49\linewidth]{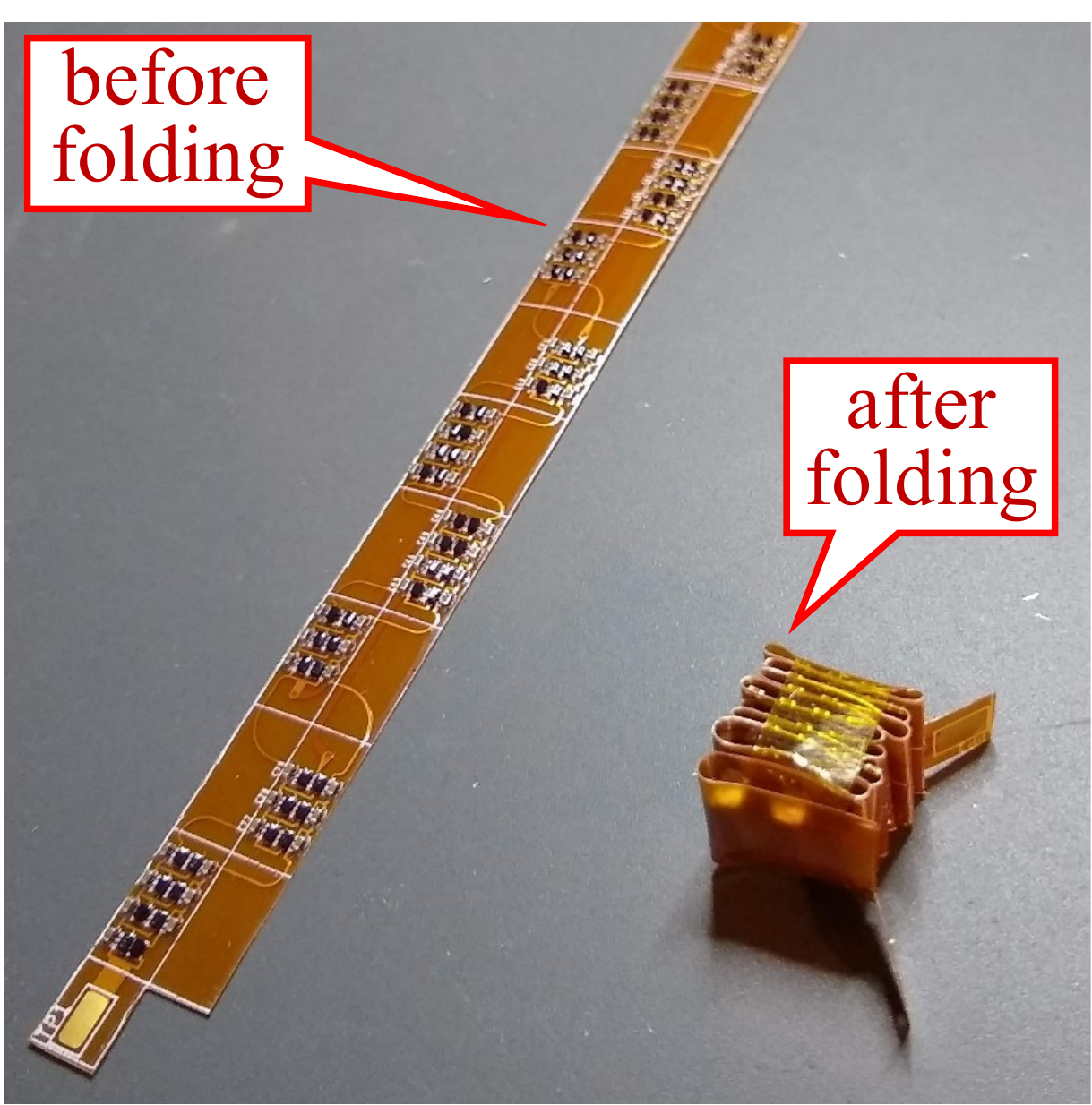}\label{fig:proto1_photo2}}
     \hfill
     \subfloat[][]{\includegraphics[width=0.49\linewidth]{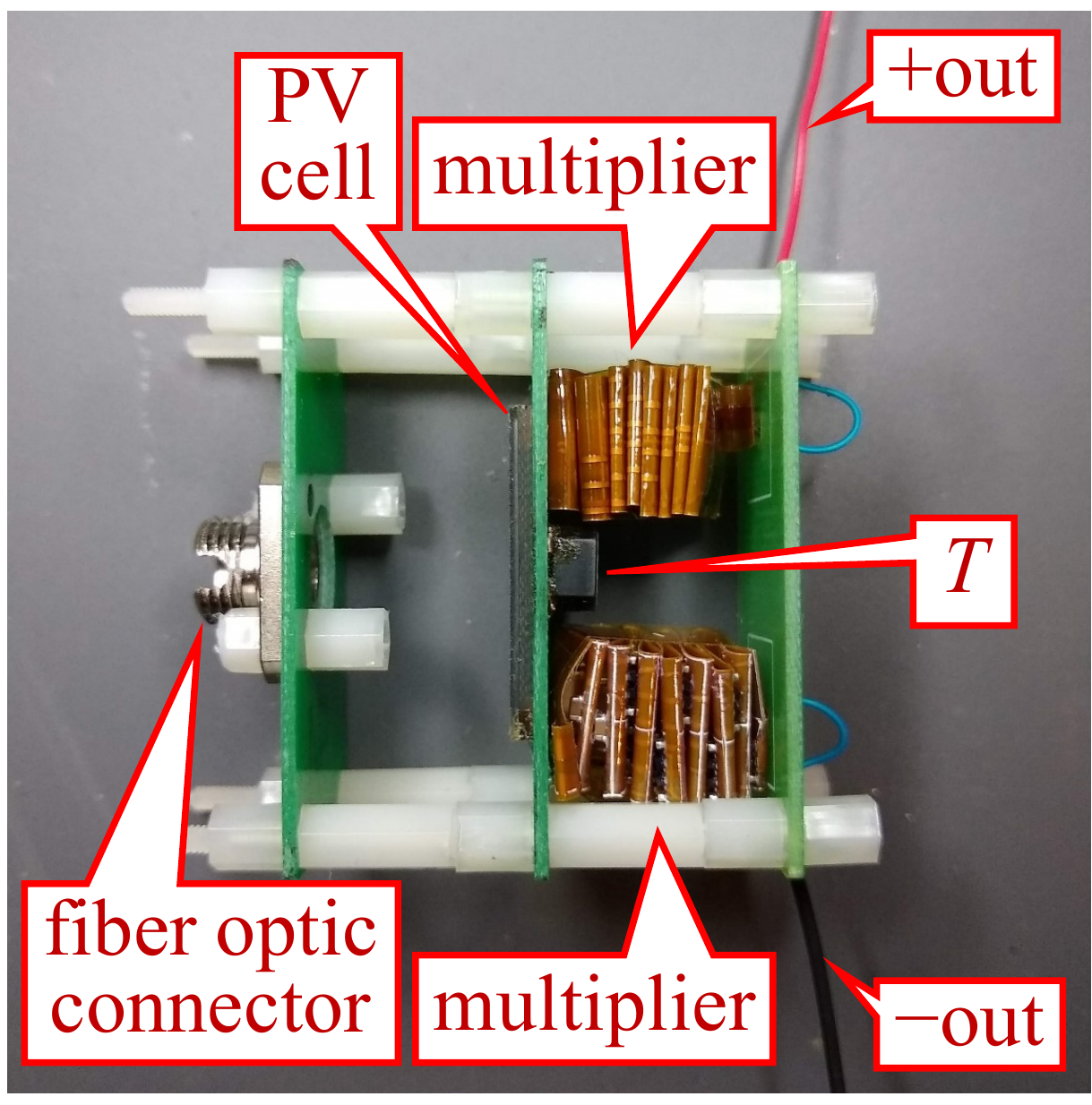}\label{fig:proto1_photo3}}
     \hfill
     \subfloat[][]{\includegraphics[width=0.49\linewidth]{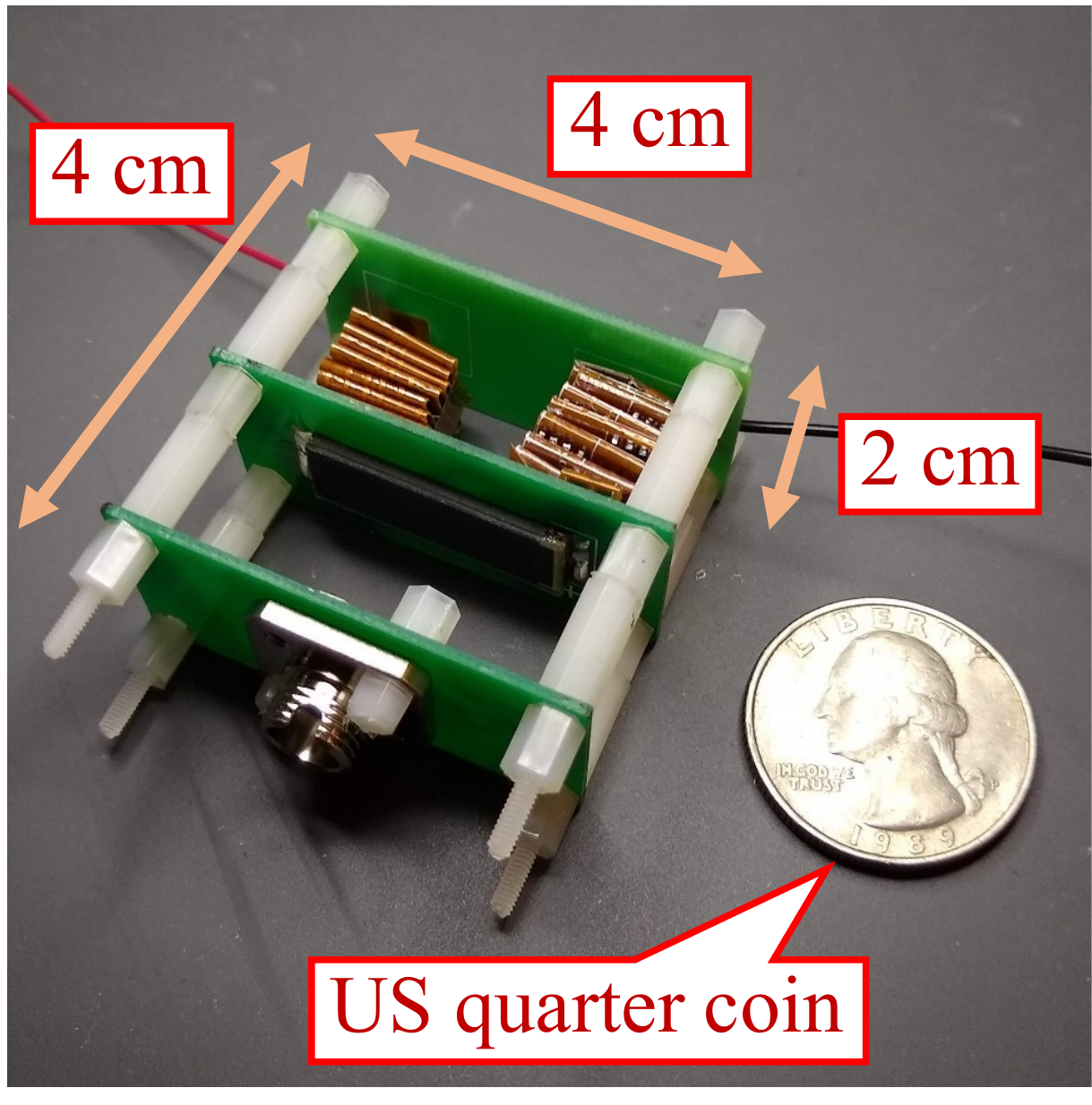}\label{fig:proto1_photo4}}
     \hfill
     \subfloat[][]{\includegraphics[width=\linewidth]{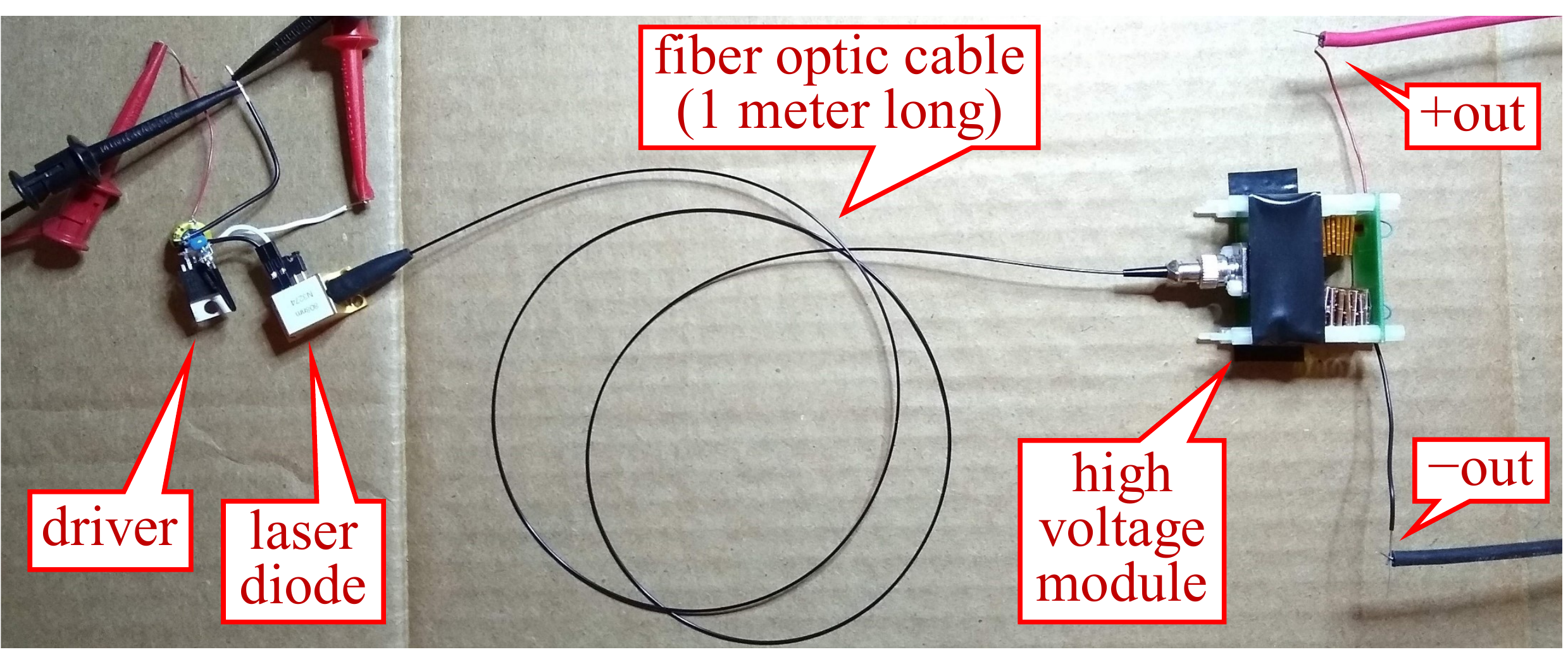}\label{fig:proto1_exp_setup}}
     \caption{The prototype high voltage dc generator powered by fiber-coupled laser. (a) Schematic. (b) Implemented assembly of laser and a PV cell. A visible laser (638~nm) is used in this photo only for the purpose of showing the laser-illuminated area. All subsequent experiments were conducted with 808~nm laser. (c) Voltage multipliers built on a flexible circuit board, before folding (left) and after (right). (d) Completed high voltage generator module after assembly. (e) Size comparison with a US quarter coin. (f) Experimental setup.}
     \label{fig:proto1}
\end{figure}

Fig.~\ref{fig:proto1} shows the implemented light-powered high voltage generator.
As described in Fig.~\ref{fig:proto1_sch}, the circuit consists of a PV cell, a 1:100 step-up transformer, and a 84-stage bipolar Cockcroft-Walton voltage multiplier.
The laser is pulsed at frequency $f$ and as a result the output voltage $V_{out}$ appears across the load resistor $R_{load}$.
The multiplier is built on a flexible circuit board as shown in Fig.~\ref{fig:proto1_photo2}.
This is so that we can fold the multiplier into a small box-shaped volume and thus realize a compact overall power supply design, as can be seen in Fig.~\ref{fig:proto1_photo3}.
Table~\ref{table:proto1_sch} lists the parts used in this implementation.

\begin{table}
    \caption{Component list of the prototype high voltage dc generator in Fig.~\ref{fig:proto1_sch}.}
    \label{table:proto1_sch}
    \centering
    \begin{tabular}{ccc}
        \toprule
        Name & Part number, Manufacturer & Description \\
        \midrule
        laser & C3743, CivilLaser & 808~nm, 2~W \\
        {\footnotesize PV cell} & KXOB25-14X1F, IXYS & monoX-Si\\
        $T$ & LPR6235-752S, Coilcraft & 1:100 turns ratio \\
        $D$ & BAS16HLP-7, Diodes Inc. & 100~V, 215~mA \\
        $C$ & {\footnotesize GRM155R62A104KE14D}, Murata &  0.1~\si{\micro}F, X5R \\
        \bottomrule
    \end{tabular}
\end{table}

\begin{figure}
     \centering
     \includegraphics[width=\linewidth]{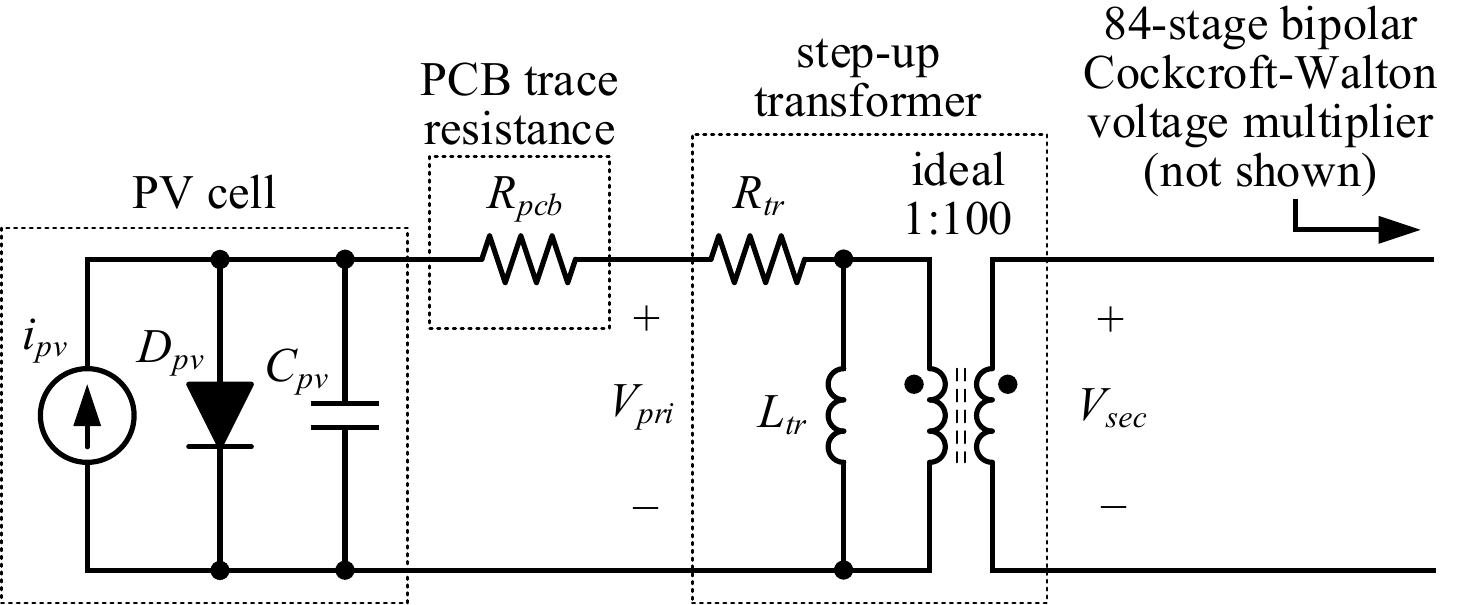}
     \caption{The equivalent circuit model of the prototype high voltage dc generator in Fig.~\ref{fig:proto1_sch}.}
     \label{fig:proto1_equiv}
\end{figure}

\blue{
The equivalent circuit model of Fig.~\ref{fig:proto1_sch} is shown in Fig.~\ref{fig:proto1_equiv}.
Pulsed laser shining on the solar cell generates a relatively small positive current $i_{pv}$.
This current by itself may not be large enough for a sufficient ac voltage amplitude $V_{pri}$.
}

\blue{
To maximize $V_{pri}$, we take advantage of the parallel $LC$ circuit consisting of the transformer's magnetizing inductance $L_{tr}$, the PV cell's junction capacitance $C_{pv}$, and the voltage multiplier's input capacitance.
We match the frequency of the pulsed laser (hence the frequency of the input current $i_{pv}$) to the resonant frequency of the $LC$ circuit.
In this way the resonant ac current dominates the total current circulating in the circuit, resulting in an ac voltage $V_{pri}$ much larger than without resonance.
}

\subsection{Equivalent circuit model and characterization}

We firstly characterize the implemented power supply in Fig.~\ref{fig:proto1_sch} using the equivalent circuit model of Fig.~\ref{fig:proto1_equiv}.
The first step is to determine the PV cell's circuit model.
The input current $i_{pv}$ is equal to 98~\% of the number of incident photons per second according to the datasheet (\cite{datasheet2019ixys}; the external quantum efficiency for 808~nm wavelength is 98~\%).
The diode $D_{pv}$ is a silicon p-n junction diode with a forward voltage of around 0.6~V.

To figure out the PV cell's capacitance $C_{pv}$, we temporarily disconnect the voltage multiplier from the transformer to make a parallel $LC$ tank.
This $LC$ tank consists of $C_{pv}$ and the transformer's magnetizing inductance $L_{tr}$ of 7.5~\si{\micro}H.
We then shine a pulsed laser light at 50~\% duty cycle on the PV cell while changing the laser's pulsing frequency.
The $LC$ tank is found to resonate at 74~kHz, from which $C_{pv}$ is calculated as 0.62~\si{\micro}F.

To measure the parasitic series resistance $(R_{pcb} + R_{tr})$, we firstly use pulsed laser to cause resonance at 74~kHz, then turn off the laser and observe the oscillating voltage across the PV cell.
The voltage decays exponentially with the time constant of 96.2~\si{\micro}s, from which $(R_{pcb} + R_{tr})$ is found to be 0.16~$\Omega$.


\subsection{Test results: single module}
\label{subsec:single_module}

We test the performance of the circuit in Fig.~\ref{fig:proto1} using a range of load resistances and laser pulsing frequencies.
Throughout the test, the laser pulse's duty cycle is fixed to 50~\% and its average output power to 1.2~W (i.e., 2.4~W peak power).

\blue{
For voltage measurements we use a high voltage differential probe, \textit{Rigol RP1100D}, which has an input resistance of 100~M$\Omega$.
This input resistance of the probe is too low for the load resistance range we want to use, from 100~M$\Omega$ up to 2100~M$\Omega$.
Connecting the probe in parallel with a load resistance would cause significant loading effect.
}

\blue{
To circumvent this issue, we connect multiple 100~M$\Omega$ resistors in series to the differential probe and use the resistor--probe--resistor chain as the load resistor for the experiment.
This resistor chain doubles as an $n$-to-1 voltage divider, which adequately scales down the voltage and prevents the possible overvoltage damage to the probe.
The use of voltage dividers is the reason for discrepancies between the apparent and actual voltage scales in the oscilloscope screen captures in the following sections.
}

Results from preliminary experiments reveal that the output voltage $V_{out}$ peaks at about 12~kHz to 20~kHz pulsing frequency $f$.
We therefore vary the frequency from 10~kHz to 22~kHz in 2-kHz increments.
The load resistance \blue{$R_{load}$} is varied from 100~M$\Omega$ to 700~M$\Omega$ so that sufficient data are obtained for estimation of the open-circuit output voltage.

\begin{figure}
     \centering
     \subfloat[][]{\includegraphics[width=0.9\linewidth]{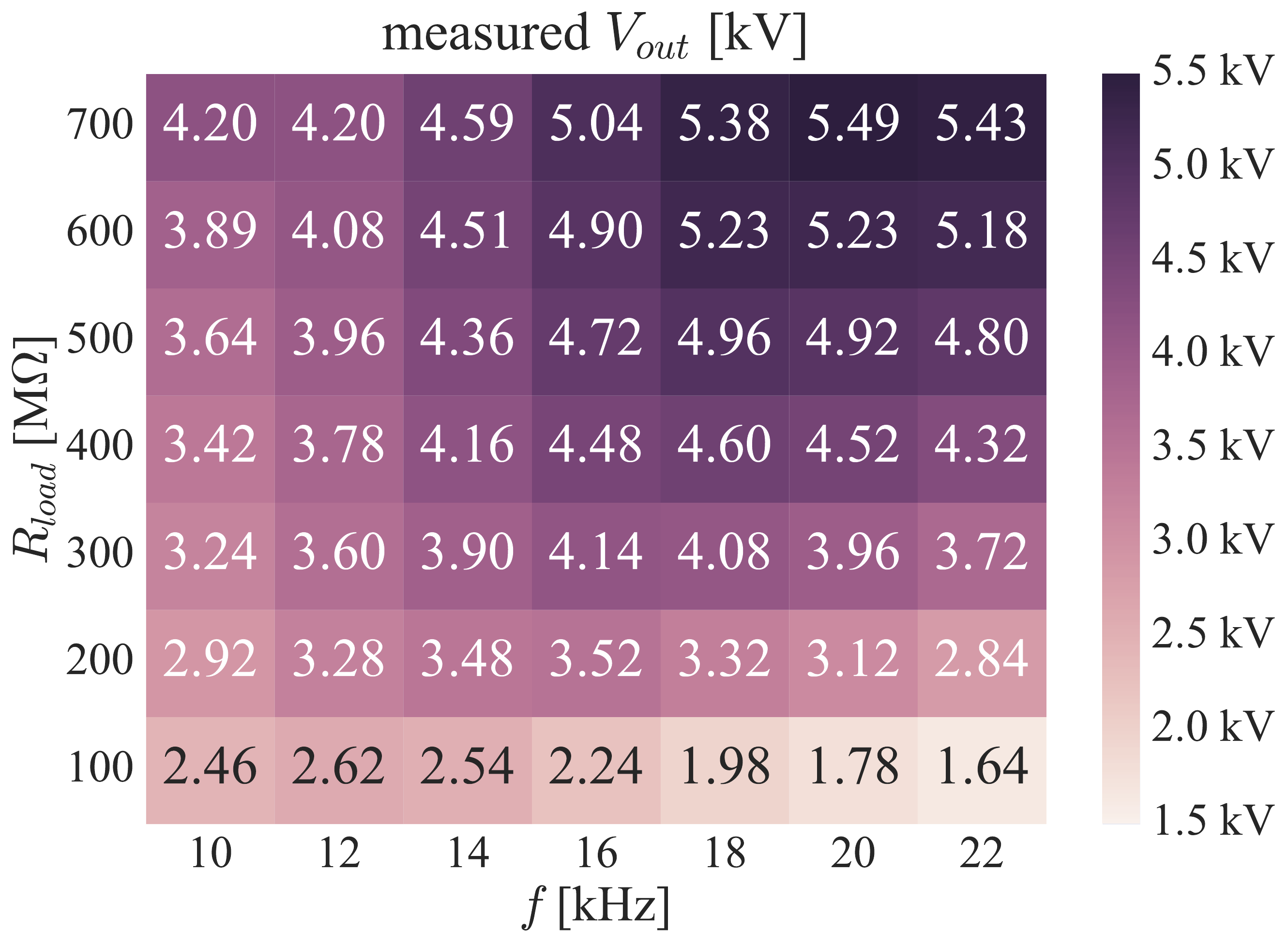}\label{fig:vout_heatmap}}
     \hfill
     \subfloat[][]{\includegraphics[width=0.7\linewidth]{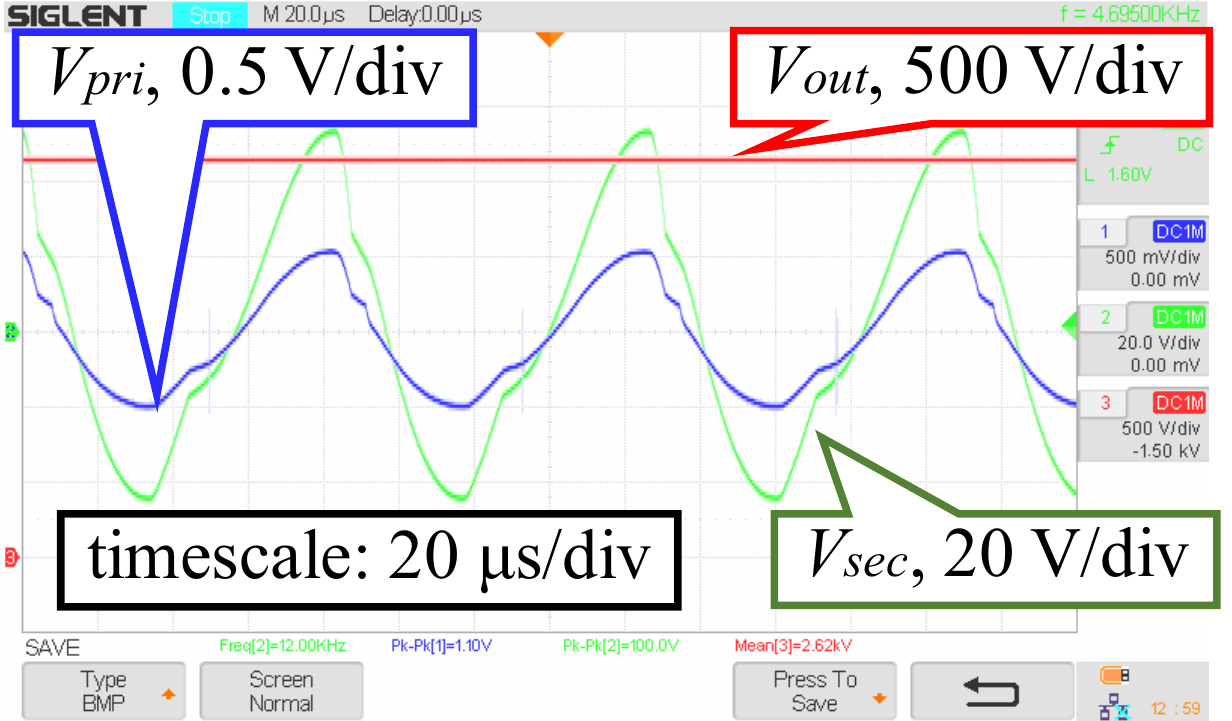}\label{fig:waveform_single_100M_12k}}
     \hfill
     \subfloat[][]{\includegraphics[width=0.7\linewidth]{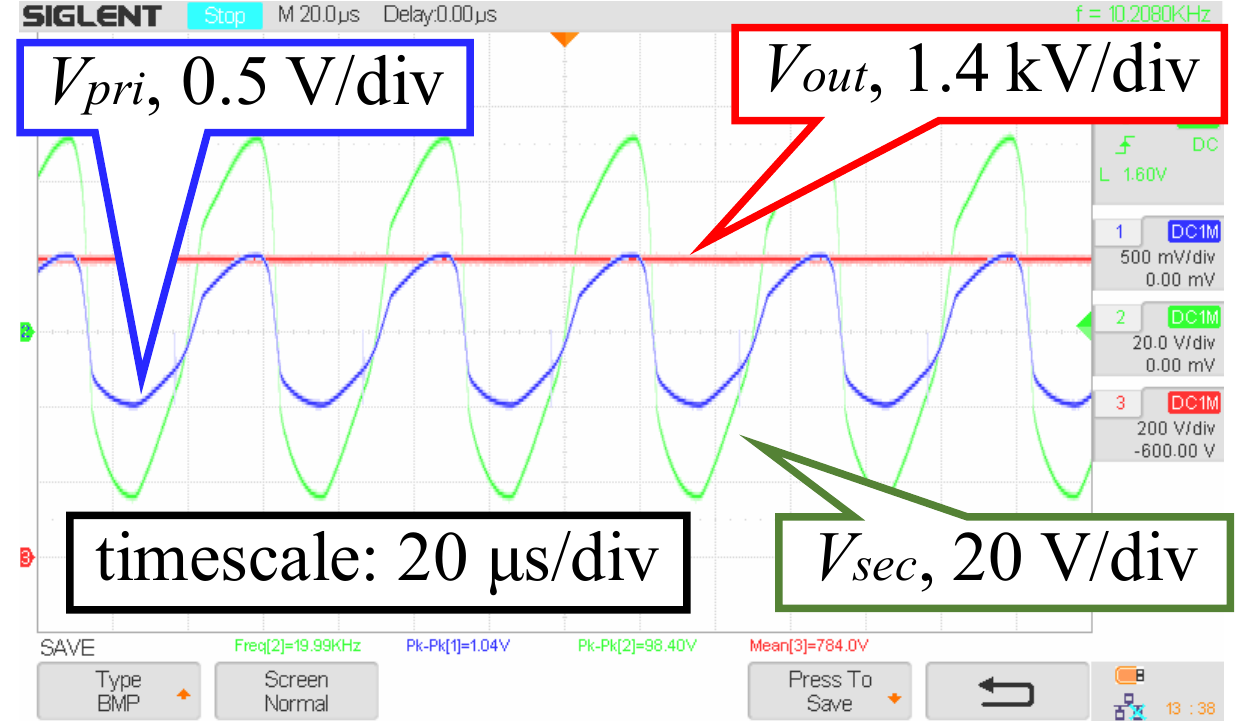}\label{fig:waveform_single_700M_20k}}
     \hfill
     \subfloat[][]{\includegraphics[width=0.9\linewidth]{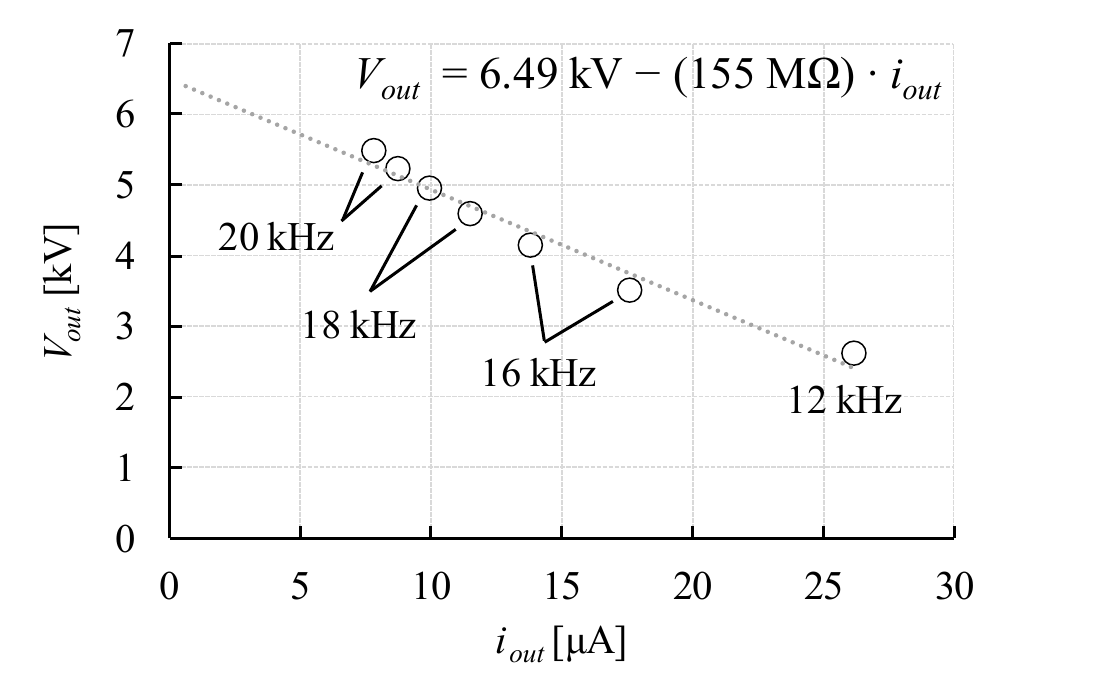}\label{fig:vi_curve}}
     \caption{Summary of the test results for a single high-voltage module. (a) Heat map of the measured $V_{out}$ values in kilovolts for the entire $R_{load}$ and $f$ ranges. (b) Waveforms at the point of highest measured output power, $R_{load}$ of 100~M$\Omega$ and $f$ of 12~kHz. (c) Waveforms at the point of highest measured output voltage, $R_{load}$ of 700~M$\Omega$ and $f$ of 20~kHz. \blue{$V_{out}$ is measured via a 7:1 voltage divider, hence the seven times discrepancy between the apparent and actual voltage scales.} (d) $I$--$V$ curve consisting of data points with the highest output voltage for each load resistance value.}
     \label{fig:data_single}
\end{figure}

Fig.~\ref{fig:data_single} summarizes the test results.
Fig.~\ref{fig:vout_heatmap} shows measured $V_{out}$ values in kilovolts for the entire $R_{load}$ and $f$ ranges.
\blue{
Two trends stand out in the plot.
First, the output voltage $V_{out}$ increases with $R_{load}$.
This is because a higher load resistance causes a low average output current, which results in a less output voltage drop as described by (\ref{equ:DeltaV}).
Second, as $R_{load}$ increases, the frequency $f$ that maximizes $V_{out}$ also increases.
This positive shift in the resonant frequency is because a higher $R_{load}$ leads to a larger $V_{out}$ as previously explained, and the larger $V_{out}$ means a larger ac voltage swing across each of the rectifying diodes.
Because a p-n junction diode capacitance decreases exponentially with the reverse bias voltage, a higher ac voltage swing means a smaller effective capacitance presented by the diodes.
As a result, the sum of all the diode junction capacitances decreases, thus the voltage multiplier's input capacitance decreases, and consequently the resonant frequency $f$ increases.
}

Within the \red{set of parameters examined,} \blue{data of Fig.~\ref{fig:vout_heatmap},} $R_{load}$ of 100~M$\Omega$ and $f$ of 12~kHz yields the highest output power 68.6~mW, resulting in the light-to-electricity power conversion efficiency of 5.7~\% (Fig.~\ref{fig:waveform_single_100M_12k}).
In addition, $R_{load}$ and $f$ of 700~M$\Omega$ and 20~kHz yields $V_{out}$ of 5.49~kV which is the highest output voltage measured from this single high-voltage module (Fig.~\ref{fig:waveform_single_700M_20k}).

When the circuit operates at the resonance, the measured \red{peak-to-peak} voltage swing of $V_{pri}$ is always limited to approximately \red{$\pm 1$~V} \blue{$\pm 0.5$~V}, and consequently the voltage swing of $V_{sec}$, about \red{$\pm 100$~V} \blue{$\pm 50$~V} throughout the experiment.
Two of the examples can be seen in Fig.~\ref{fig:waveform_single_100M_12k} and Fig.~\ref{fig:waveform_single_700M_20k}.
The reason is probably that the positive voltage swing of $V_{pri}$ is clipped by the PV cell's intrinsic diode $D_{pv}$.

Selecting a data point of the highest output voltage for each load resistance gives the $I$--$V$ curve of Fig.~\ref{fig:vi_curve}.
From the fitted curve the open-circuit voltage is estimated to be 6.49~kV although its actual value is likely to be higher considering the rate of increase in $V_{out}$ at low values of the output current $i_{out}$.
The measured output resistance of 155~M$\Omega$ is roughly in line with theoretical calculations~\cite{kuffel2000high,Naidu2010,Ray2013}.

\subsection{Test results: three modules cascaded}

\begin{figure}
     \centering
     \subfloat[][]{\includegraphics[width=\linewidth]{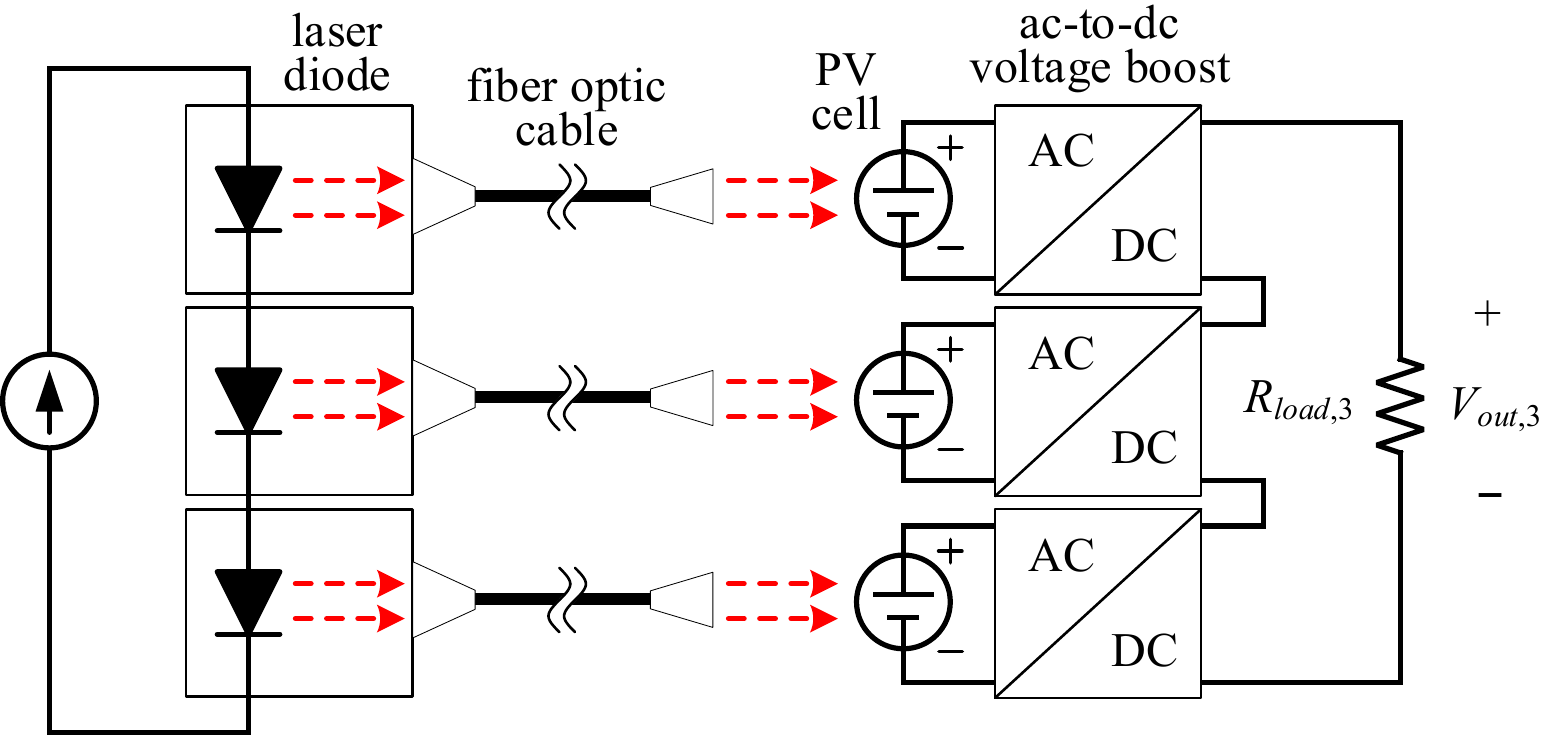}\label{fig:proto2_sch}}
     \hfill
     \subfloat[][]{\includegraphics[width=0.33\linewidth]{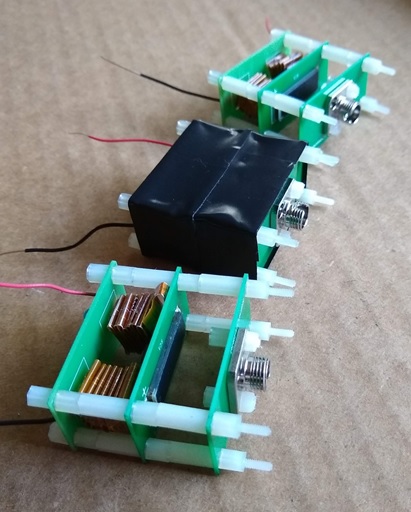}\label{fig:proto2_photo1}}
     \hfill
     \subfloat[][]{\includegraphics[width=0.33\linewidth]{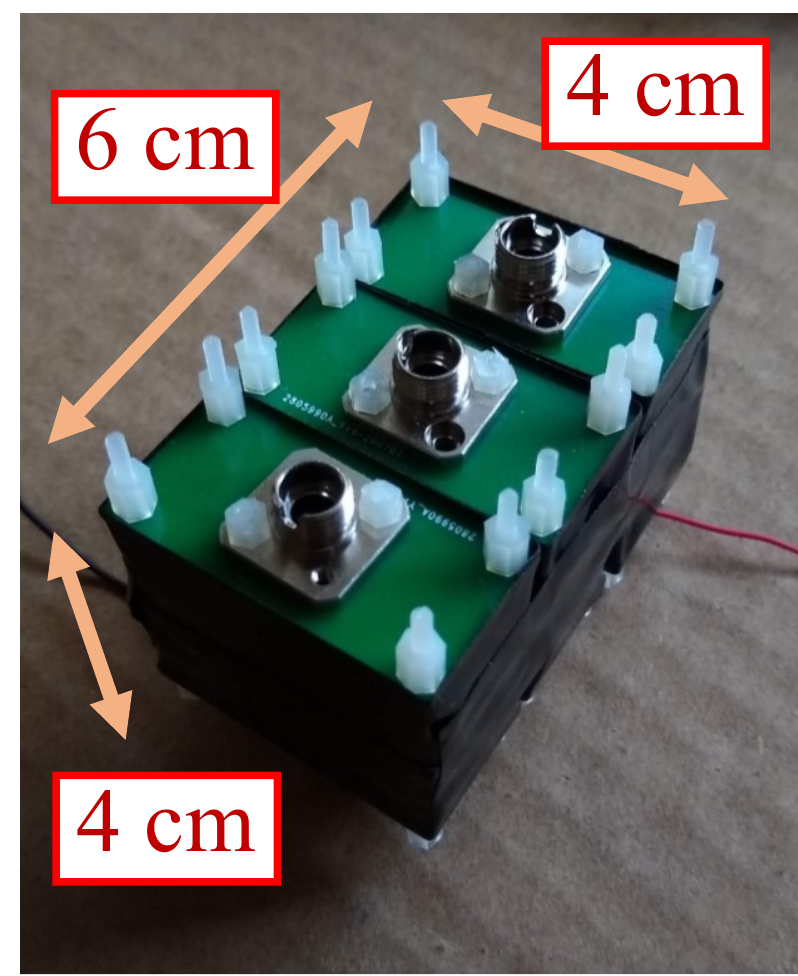}\label{fig:proto3_photo2}}
     \hfill
     \subfloat[][]{\includegraphics[width=0.33\linewidth]{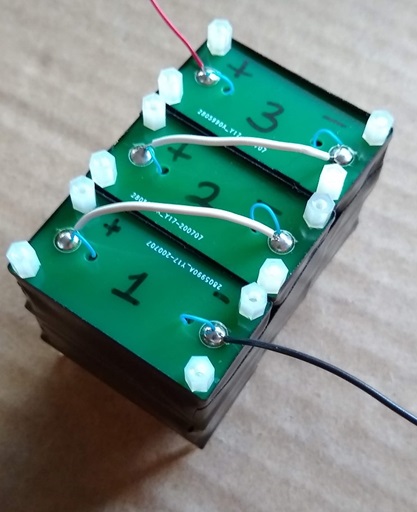}\label{fig:proto3_photo3}}
     \hfill
     \subfloat[][]{\includegraphics[width=\linewidth]{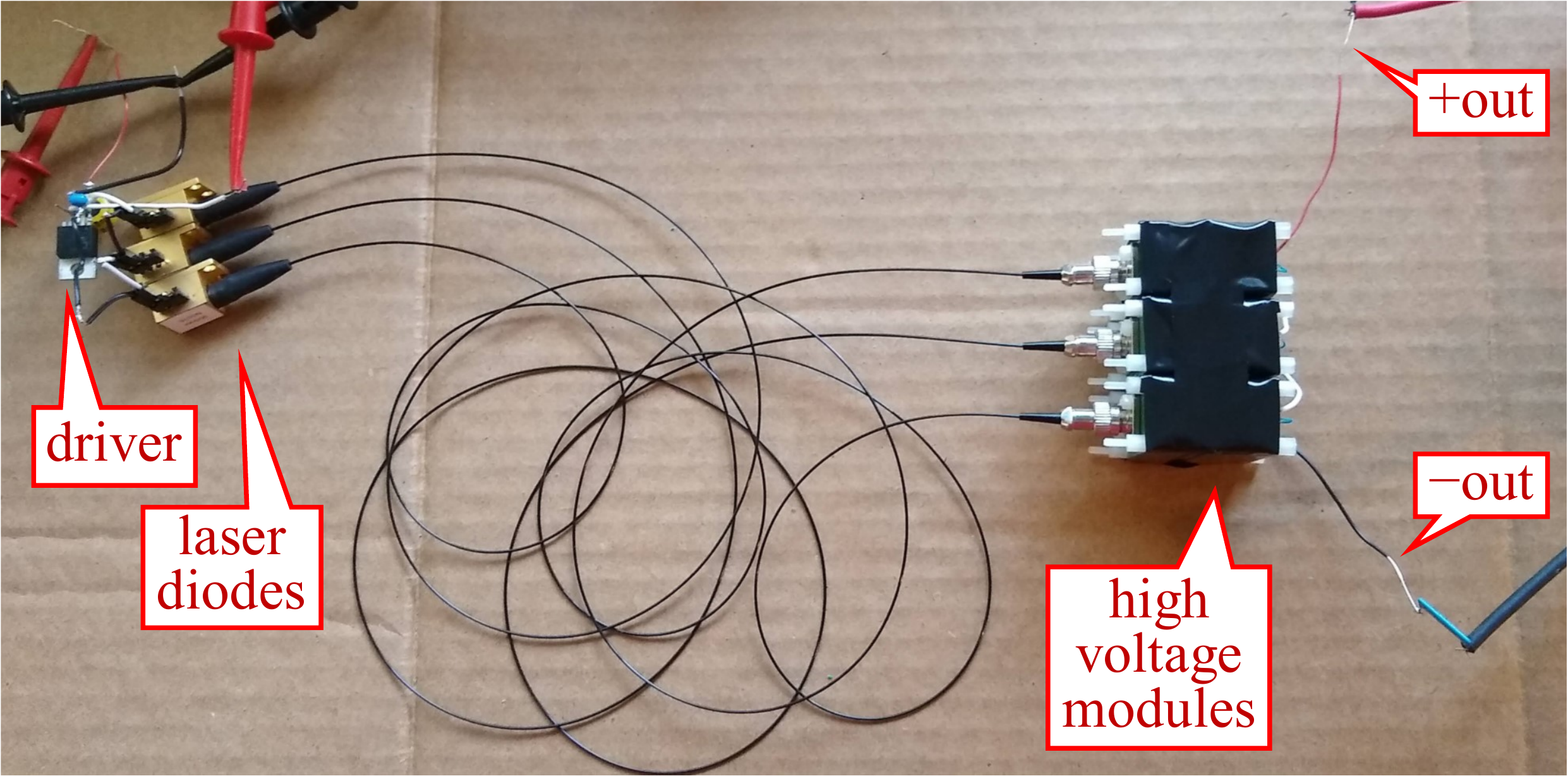}\label{fig:proto2_exp_setup}}
     \caption{The prototype laser-powered high voltage dc generator consisting of three modules cascaded. (a) Schematic. (b) High-voltage modules, before stacking. (c) Finished high-voltage modules (front). (d) Finished high-voltage modules (back). (e) Experimental setup.}
     \label{fig:proto2}
\end{figure}

We proceed to cascade three high-voltage modules in series and thereby produce an output voltage three times that of a single module.
The purpose is to demonstrate that cascading multiple units of the proposed circuit allows one to obtain a higher output voltage without performance degradation usually seen in other high voltage generation methods.
Fig.~\ref{fig:proto2} shows the implemented circuit and the experimental setup.

\begin{figure}
     \centering
     \subfloat[][]{\includegraphics[width=0.9\linewidth]{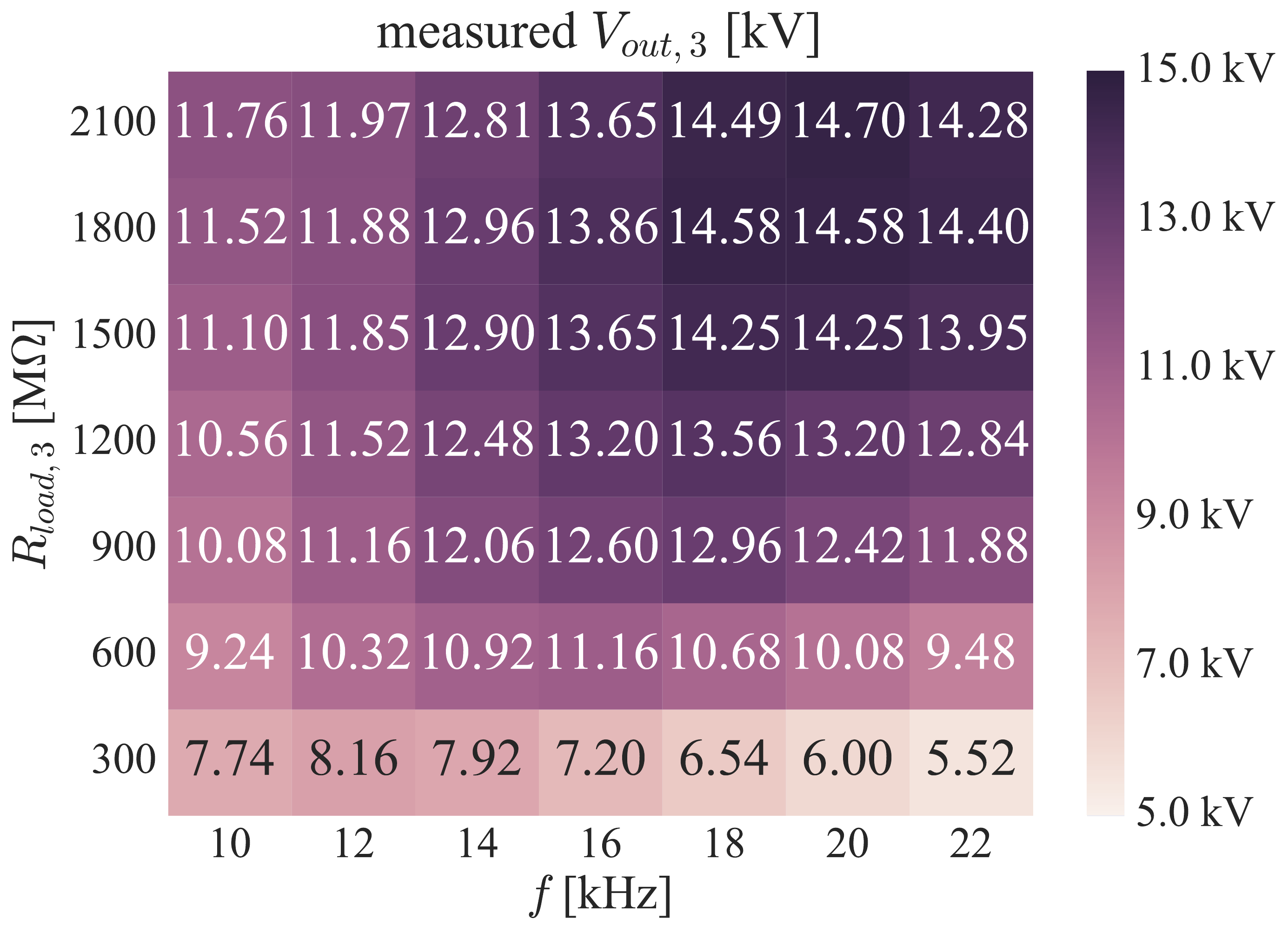}\label{fig:vout_heatmap_3module}}
     \hfill
     \subfloat[][]{\includegraphics[width=0.9\linewidth]{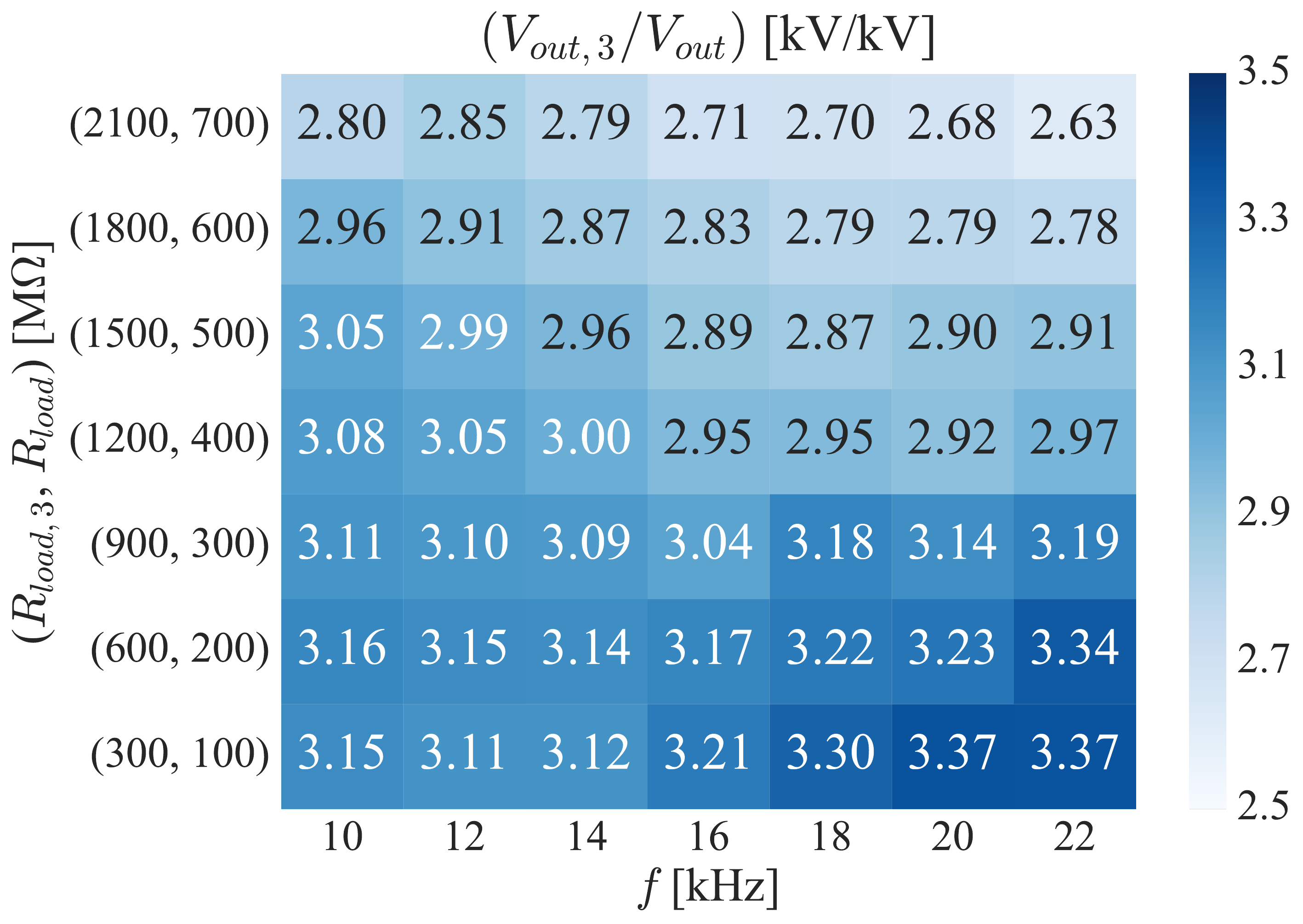}\label{fig:comp_3vs1_heatmap}}
     \hfill
     \subfloat[][]{\includegraphics[width=0.85\linewidth]{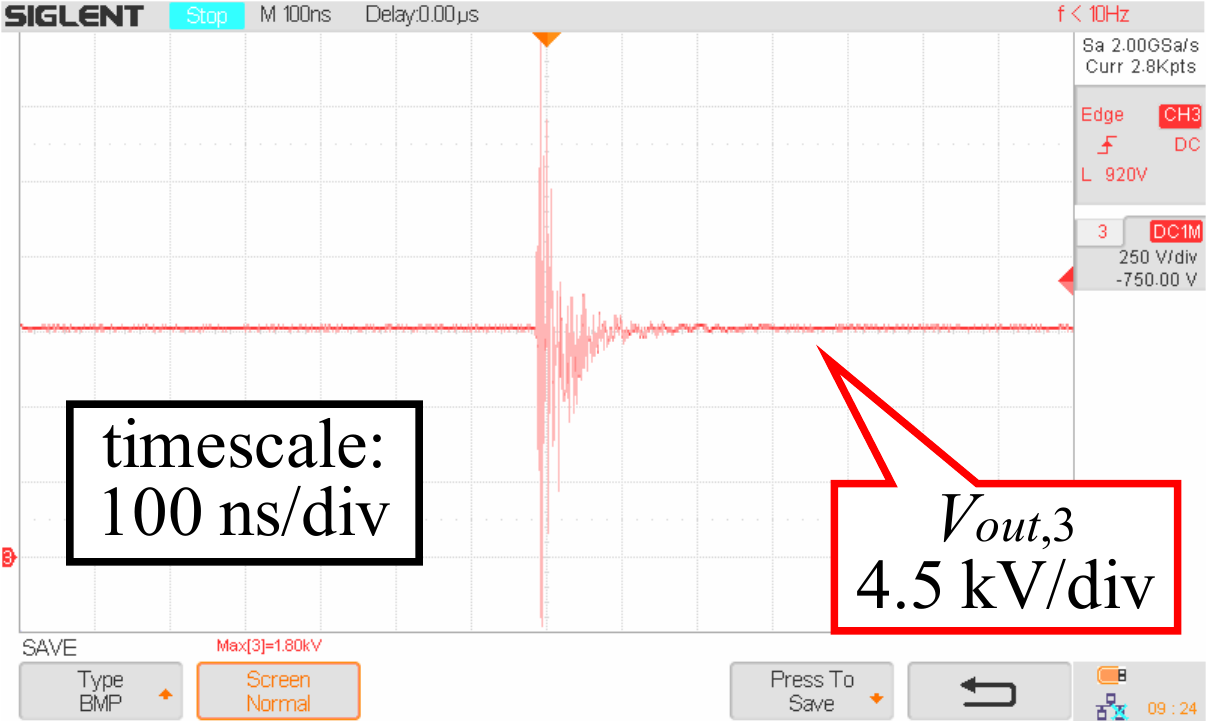}\label{fig:vout_waveform_3module}}
     \caption{Test results for three high-voltage modules cascaded. (a) Heat map of the measured high voltage output $V_{out,3}$ in kilovolts. (b) The ratio of three-module output voltage $V_{out,3}$ to the single module output voltage $V_{out}$. (c) One of the occasional voltage spikes measured when $V_{out,3}$ was larger than 12~kV. \blue{$V_{out}$ is measured via a 18:1 voltage divider, hence the 18 times discrepancy between the apparent and actual voltage scales.}}
     \label{fig:results_3module}
\end{figure}

Fig.~\ref{fig:vout_heatmap_3module} shows the test results.
Fig.~\ref{fig:vout_heatmap_3module} is the heat map of the measured output voltage $V_{out,3}$.
As one can see from Fig.~\ref{fig:comp_3vs1_heatmap} plotting $(V_{out,3}/V_{out})$, for all frequency and load conditions $V_{out,3}$ values have almost tripled compared to the output voltages $V_{out}$ from a single module shown in Fig.~\ref{fig:vout_heatmap}.
The voltage increase is slightly more than three-fold at a low voltage ($V_{out,3} \lessapprox 12$~kV) because newly built modules produced a higher voltage then the one tested in section~\ref{subsec:single_module}.
On the other hand, the voltage increase is reduced to less than three-fold as the output voltage becomes higher ($V_{out,3} \gtrapprox 12$~kV), presumably because of corona discharge that onsets at around 6~kV ground-referenced potential.
To corroborate our claim, a hissing sound from the circuit was evident when the output voltage exceeded 10~kV.
Also, occasional voltage spikes such as Fig.~\ref{fig:vout_waveform_3module} were continually measured at the output, suggesting partial discharges (localized electrical discharge) occurring within the circuit.


\subsection{Potential improvements and future opportunities}

In designing and implementing the light-powered high voltage generator, we encountered several practical difficulties.
First, there was no off-the-shelf PV cell readily available to match the area irradiated by the laser.
As can be seen in Fig.~\ref{fig:proto1_photo1}, the PV cell we used is in a long rectangular shape whereas the laser-illuminated area is only a small circular patch at the center.
As a result, more than two third of the PV cell junction area contributes nothing but instead increases the capacitance $C_{pv}$ and the current through $D_{pv}$, both of which increases the power loss.
Second, although using a PV array instead of a single cell to increase the ac voltage swing would have helped, such design was not attempted due to the difficulty of evenly distributing the laser light over multiple cells.
Third, related to first and second, low voltage swing from the PV cell led to the use of a high turns-ratio transformer.
Because of this high turns ratio, the capacitance of the voltage multiplier presented a large value when reflected to the transformer's primary side.
This effect, combined with the presence of extra large $C_{pv}$, lowered the resonant frequency of the circuit and thus resulted in the huge output resistance as seen in Fig.~\ref{fig:vi_curve}~\cite{kuffel2000high,Naidu2010,Ray2013}.

In view of our findings, we list the following possible modifications to improve the performance of the circuit and bring it closer to becoming a competitive solution for high voltage applications.
The first is to customize the PV cell's shape so that the unused junction area is minimized for a smaller capacitance $C_{pv}$.
The second is to use a multi-cell array with a means of distributing light so that a high-amplitude ac voltage is produced.
Then the transformer turns ratio can be lowered.
The first and the second approaches combined will increase the resonant frequency of the circuit and, therefore, increase the available output power from the multiplier.
Needless to say, using a higher-power light source with the wavelength closer to the absorption edge of the PV cell will further improve the performance of the circuit.


\section{\blue{Application Example on Electroadhesion}}
\label{sec:app_example}

\blue{This section shows a possible application of the proposed technique.
We demonstrate a simple gripper that grips and moves an object by using an electroadhesive pad~\cite{monkman1989principles,monkman2003electroadhesive} powered by the developed high voltage generator.
}

\begin{figure}
     \centering
     \subfloat[][]{\includegraphics[width=0.49\linewidth]{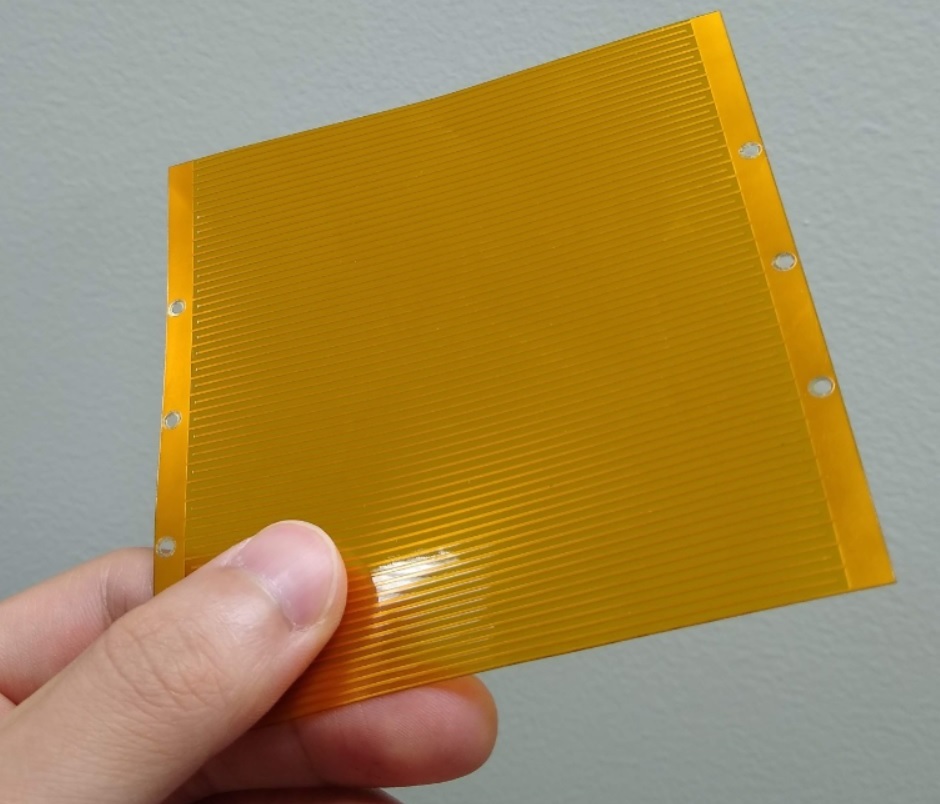}\label{fig:photo_ead_pad}}
     \hfill
     \subfloat[][]{\includegraphics[width=0.49\linewidth]{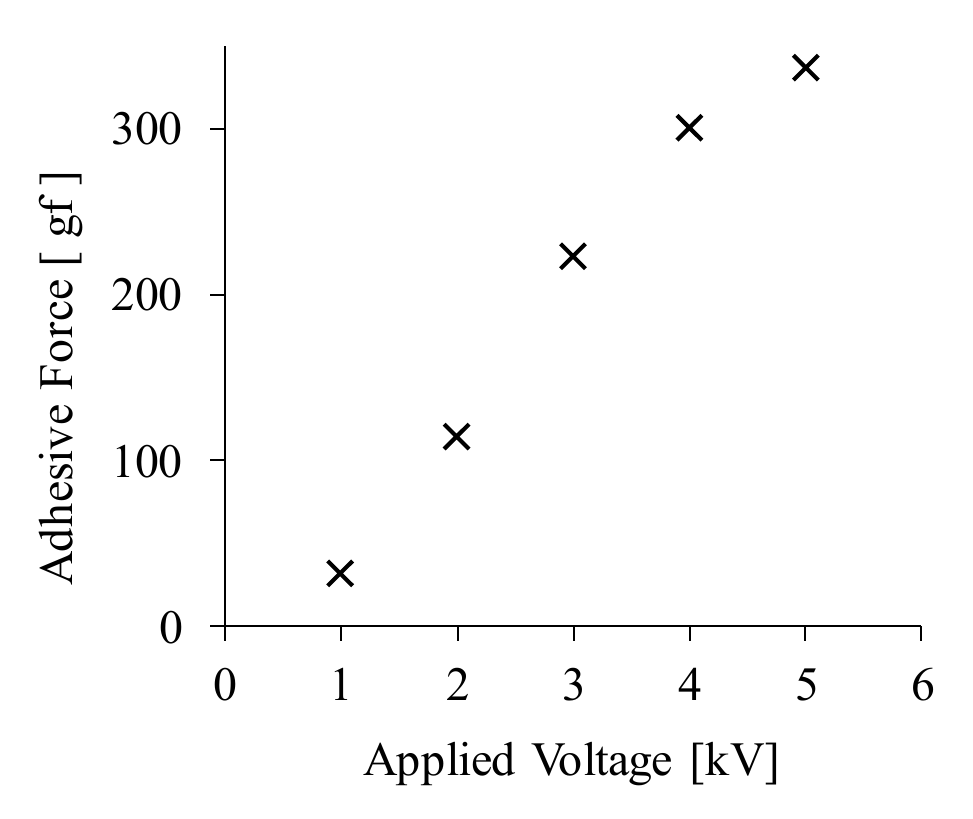}\label{fig:plot_VvsF}}
     \caption{\blue{The electroadhesive pad used for the demonstration, measuring 10~cm$\times$10~cm in size. (a) Photograph. (b) Adhesive force versus applied voltage measured under an ideal condition (taken from our previous work~\cite{9113652}).}}
     \label{fig:ead_pad}
\end{figure}

\blue{
The electroadhesive pad, shown in Fig.~\ref{fig:photo_ead_pad}, consists of interdigitated electrodes laid on and insulated by a thin polyamide film.
When a voltage of several kilovolts is applied to the electrodes while the pad is in contact with a material surface, an opposite static charge is induced on the surface.
The resulting electrostatic attraction between the pad and the material enables gripping and material handling.
Particularly in the case of our electroadhesive pad, the adhesive force under an ideal condition is measured to be over 100~gram-force at the applied voltage of over 2~kV, as plotted in Fig.~\ref{fig:plot_VvsF}.
(The data are taken from our previous work in which we used the same pad for a similar type of demonstration~\cite{9113652}.)
}

\blue{
Let us assume a scenario where one wants to design a gripper that has an electroadhesive pad mounted on tip of an actuation mechanism such as a robot arm.
In this scenario, the designer wants to electrically isolate the high voltage generator and the pad so that no single-point insulation failure leads to dangerous electrical discharge through either the user or the handled object.
This requirement for electrical safety is comparable to the Class-II standard (also known as \textit{double-insulation}) determined for consumer appliances by International Electrotechnical Commission (IEC)~\cite{iec2016iec}.
}

\begin{figure}
     \centering
     \subfloat[][]{\includegraphics[width=0.85\linewidth]{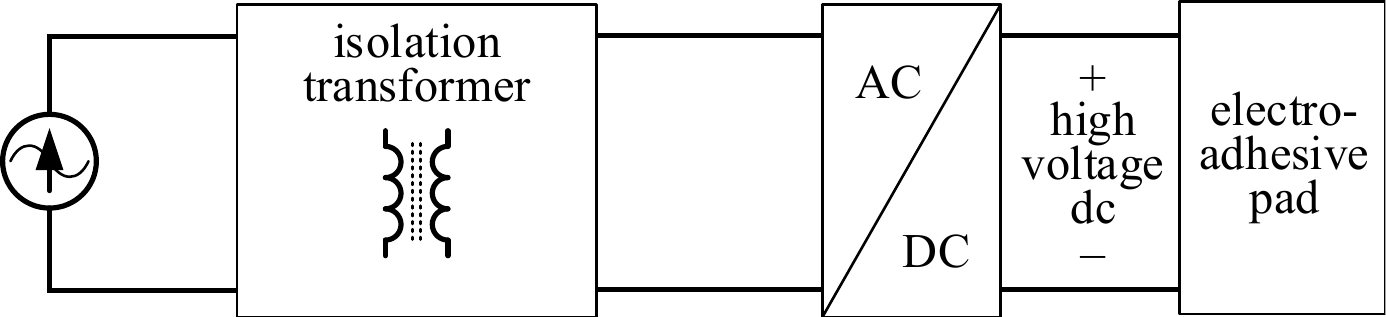}\label{fig:sch_tx_approach}}
     \hfill
     \subfloat[][]{\includegraphics[width=0.85\linewidth]{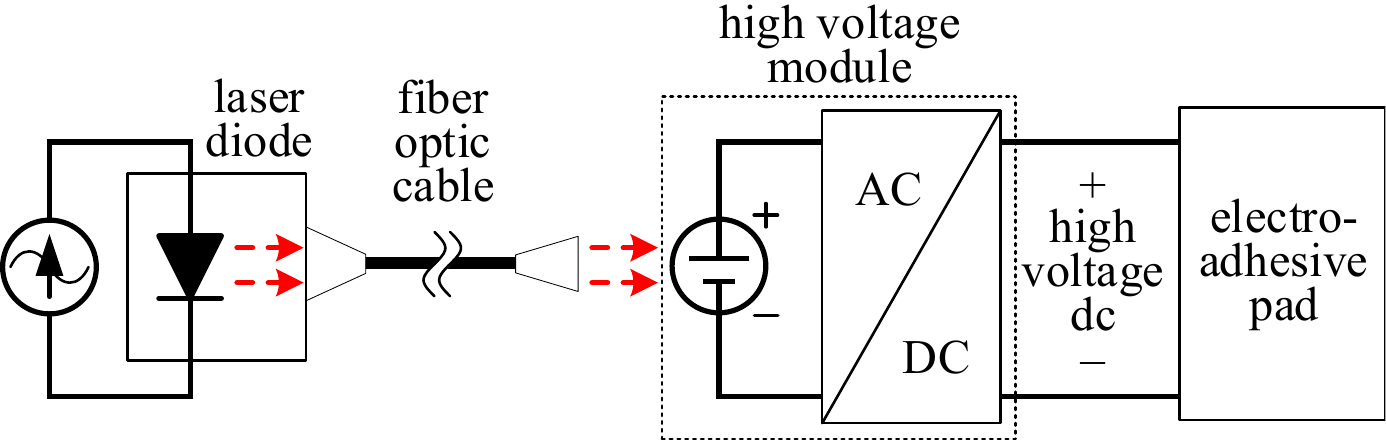}\label{fig:sch_optical_approach}}
     \caption{\blue{Structure comparison of electroadhesive gripper designs. (a) Approach using an isolation transformer. (b) Approach using the proposed high voltage generator.}}
     \label{fig:sch_approaches}
\end{figure}

\blue{
For electrical isolation, the most common and straightforward approach is to use an isolation transformer as in Fig.~\ref{fig:sch_tx_approach}.
The transformer should not only be able to withstand several kilovolts of dc potential difference between the primary and secondary windings, but also should have a low inter-winding capacitance to prevent the ac component of the discharge current from flowing across the windings.
}

\blue{
The potentially large size and weight of the isolation transformer can be avoided by using the proposed light-powered high voltage generator as described in Fig.~\ref{fig:sch_optical_approach}.
The fiber optic cable provides a mechanically flexible power coupling with minimal stray capacitance and high insulation voltage.
This scenario is where the proposed high voltage generator can be superior over conventional solutions, because the power consumption of the electroadhesive pad is minuscule (Measured current is well below 10~nA at 3~kV applied voltage.) and the means of electrical insulation dominates the total system build cost.
}

\begin{figure}
     \centering
     \includegraphics[width=0.8\linewidth]{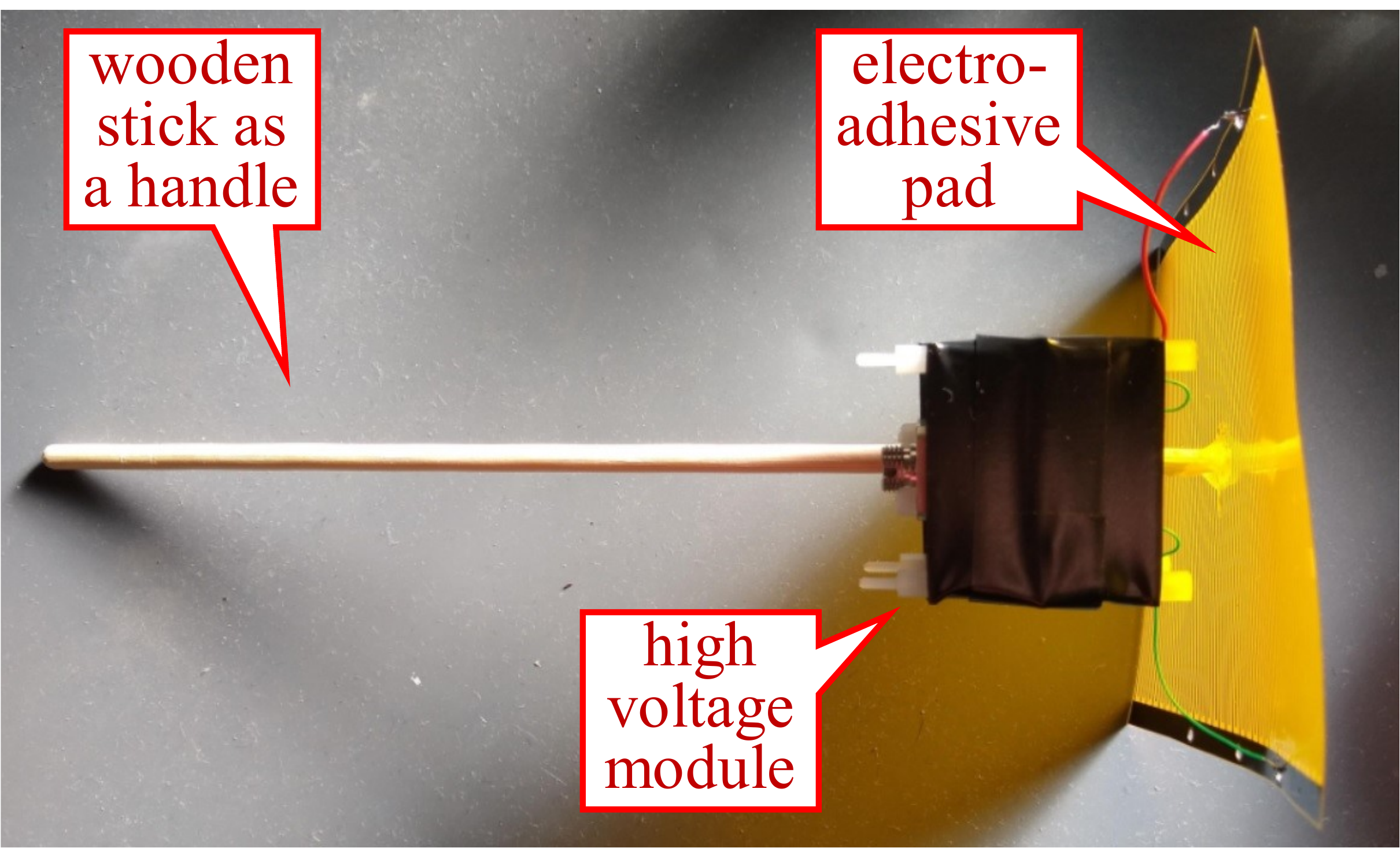}
     \caption{\blue{The implemented electroadhesive gripper.}}
     \label{fig:ead_gripper}
\end{figure}

\blue{
Fig.~\ref{fig:ead_gripper} is the implemented gripper.
We attach a wooden stick to the electroadhesive pad perpendicularly to use it as a handle.
The high voltage module is taped to the wooden stick, and the output nodes are wired to the pad's electrodes.
}

\begin{figure}
     \centering
     \subfloat[][]{\includegraphics[width=0.33\linewidth]{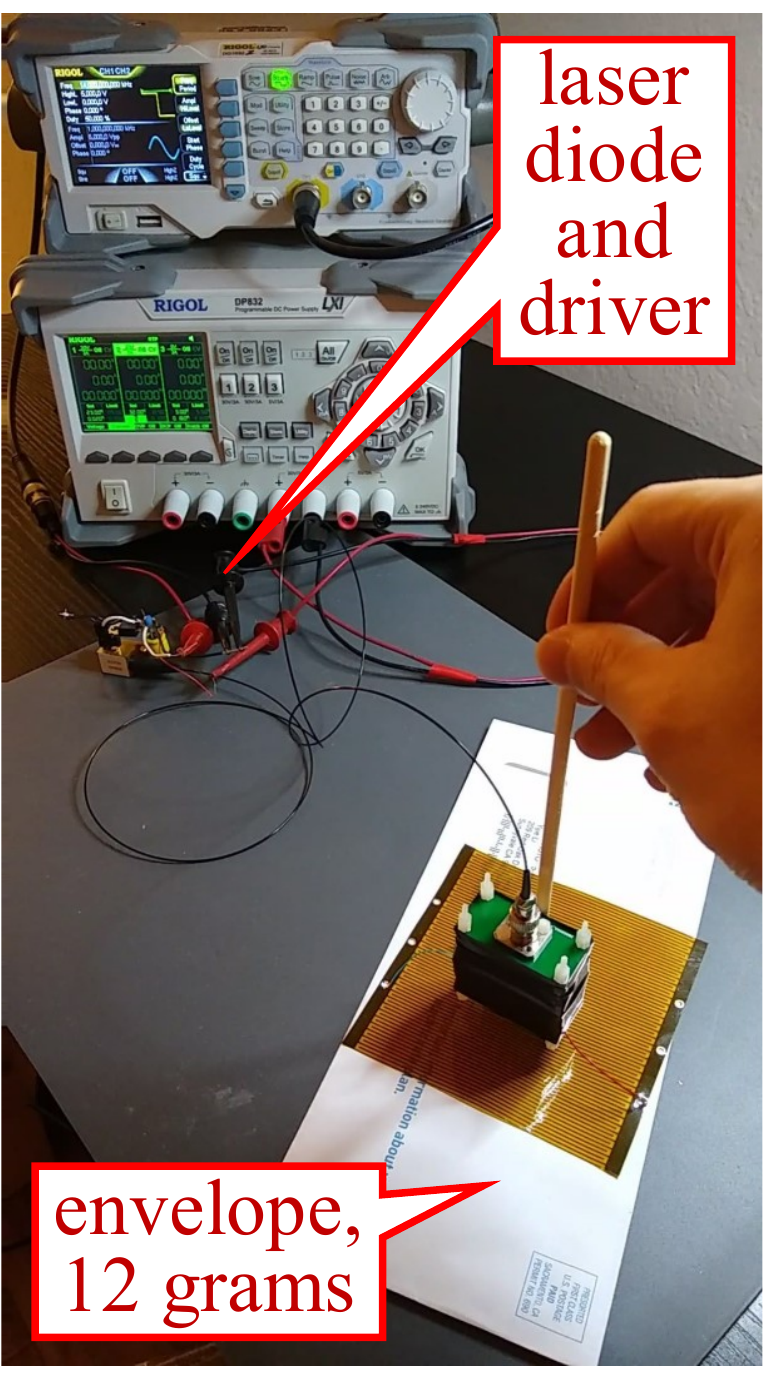}\label{fig:demo1}}
     \hfill
     \subfloat[][]{\includegraphics[width=0.33\linewidth]{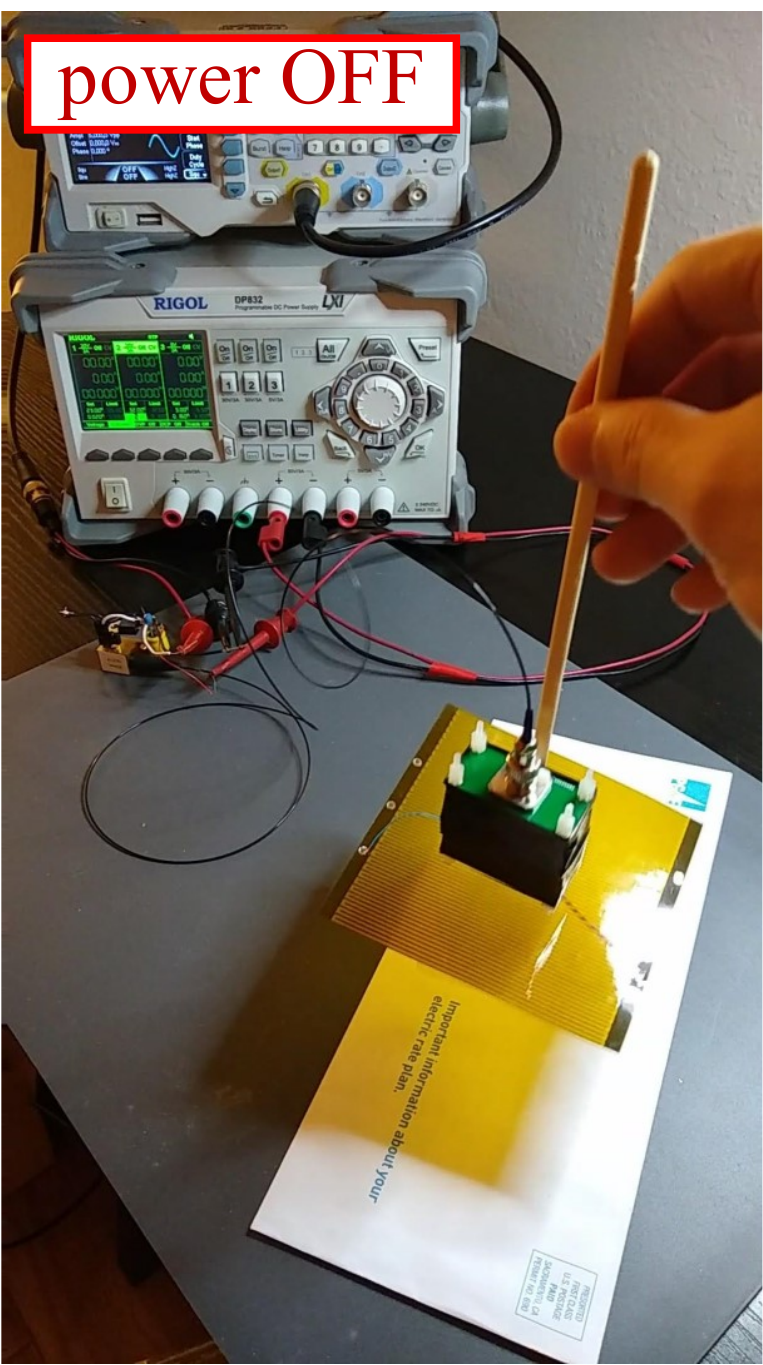}\label{fig:demo2}}
     \hfill
     \subfloat[][]{\includegraphics[width=0.33\linewidth]{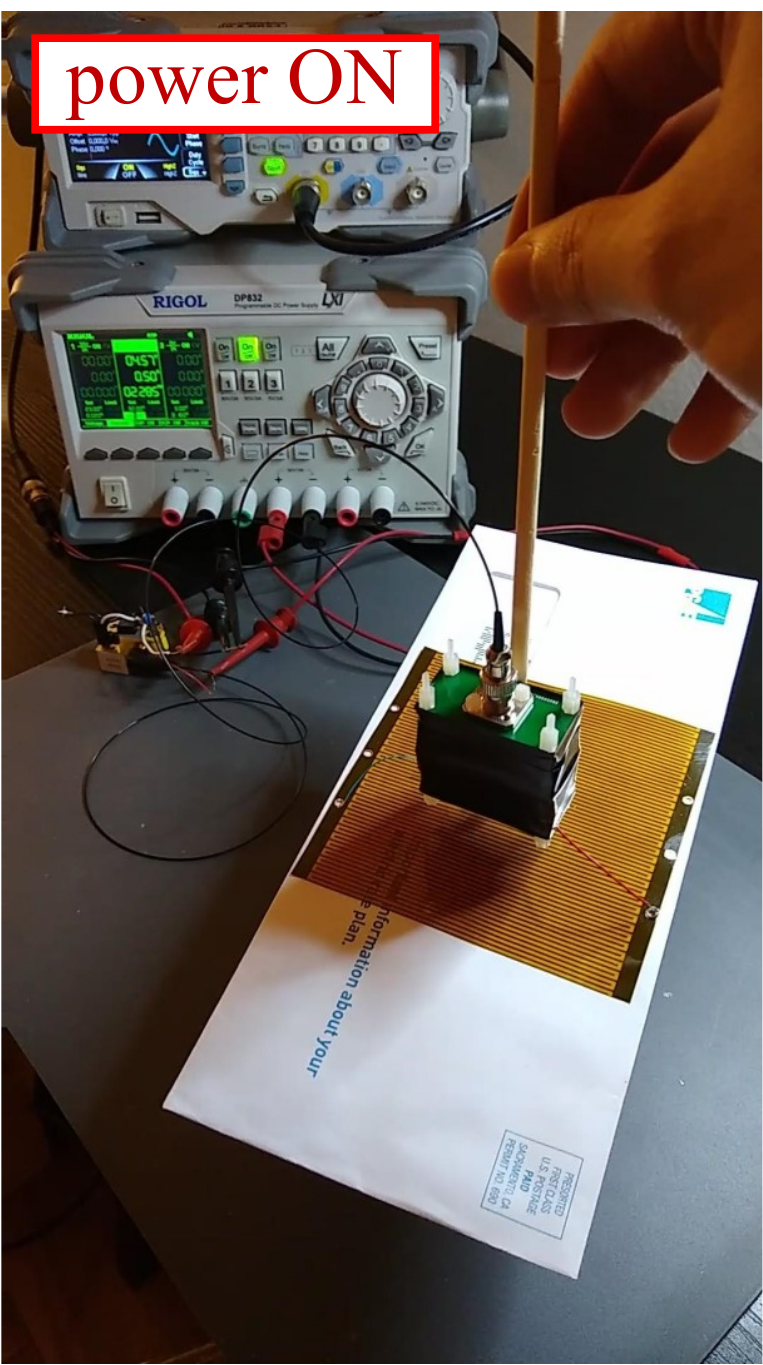}\label{fig:demo3}}
     \caption{\blue{Demonstration of the electroadhesive gripper lifting an envelope weighing 12~grams. (a) Experimental setup. (b) Gripper lifted when the laser diode is powered off. (c) Gripper lifted when the laser diode is powered on. Notice the envelope attached to the pad and lifted.}}
     \label{fig:demo}
\end{figure}

\blue{
Fig.~\ref{fig:demo} demonstrates the operation of the electroadhesive gripper.
The pad on the gripper is pressed against a regular postal envelope weighing 12~grams.
When the laser diode remains powered off (Fig.~\ref{fig:demo2}), the gripper does not lift the envelope.
When the diode is turned on (Fig.~\ref{fig:demo3}), the high voltage generator energizes the electroadhesive pad, inducing electrostatic force on the pad and lifting the envelope.
All the while, the fiber optic cable mechanically decouples the laser diode from the gripper, thereby minimally restricting the movement of the gripper.
}

\section{Conclusion}
\label{sec:conc}


In this paper, we investigated the feasibility of high voltage generation by optical power transfer.
We reviewed the existing approaches for high voltage dc generation and classified them by how the input-to-output dc isolation is achieved.
From that perspective, electrostatic machines are based on mechanical input-to-output energy coupling, and voltage multipliers are based on capacitive and/or inductive coupling.
The approach subject to our study was optical energy coupling, that is, using fiber-coupled laser and a PV cell to deliver power from the low-voltage input side to the high-voltage output side.
The optical scheme's high input-to-output electrical insulation and spatial separation are helpful for cascading multiple circuits in series for a higher output voltage.
Different from other optically powered high voltage generators using continuous-wave light and an integrated PV array, our design used pulsed laser and discrete components for simple construction.


We experimentally demonstrated the operation of the proposed pulsed-laser powered high voltage generator.
A PV cell converts a laser pulse train to a low-voltage ac voltage, which is amplified to become a high-voltage ac via a step-up transformer and finally becomes a high-voltage dc via a diode-capacitor voltage multiplier.
We used a light source of 1.2~W average output at 10-to-22~kHz pulsing frequencies to power the circuit and measured 5.5~kV dc output from a single module.
From the subsequent experiment we powered three cascaded modules using three laser diodes and measured 14.7~kV at the output.



As a concluding remark, we list a few questions to suggest future directions for improved light-based high voltage generation suitable for real-world applications.
First, \textit{higher switching frequencies}: For applications that demand a low ripple output, can a light-powered circuit be built to operate at high-enough frequencies so as to make a separate output filter capacitor unnecessary?
Second, \textit{ripple-free dc generation}: Can a ripple-free high voltage dc be obtained by massively integrated PV arrays with no switching elements?
Third, \textit{system optimization}: For a given dc output voltage, how can one determine the optimal proportion of voltage gains by a PV array, a transformer, and a multiplier?
Lastly, \textit{system miniaturization}: Other than the transformer, can all the components in the high voltage module be integrated into a monolithic circuit for compact system arrangement?
By addressing those questions we hope that future studies will identify design spaces where light-based systems are a superior solution in terms of performance, build cost, and reliability.


\bibliographystyle{IEEEtran}
\bibliography{my_bibliography.bib}

\end{document}